\documentclass[12pt]{iopart}

\usepackage{iopams} 
\usepackage{graphicx,epstopdf} 
\usepackage{caption}
\usepackage{subcaption}
\usepackage{framed}
\usepackage{comment}
\usepackage{soul}
\begin{document}

\title[]{Emergent soliton-like solutions in the parametrically driven 1-D nonlinear Schr\"odinger equation}

\author{K Dileep and S Murugesh}

\address{Department of Physics, Indian Institute of Space Science and Technology, Thiruvananthapuram 695~547, India}
\ead{dileepk.17@res.iist.ac.in}
\vspace{10pt}
\begin{indented}
\item[]
\end{indented}

\begin{abstract}
We numerically investigate the long time dynamics of spatially periodic breather solutions of the 1-D nonlinear Schr\"odinger equation under parametric forcing of the form $f(x)=f_0 \exp(iKx)$ along with dissipation. In the absence of dissipation, robust soliton-like excitations are observed that travel with constant amplitude and velocity. With dissipation, these solitons lose energy (and amplitude) yet gain speed - a characteristic not observed in an ordinary soliton. Moreover, these novel solitons are found to be stable against random perturbations.
\end{abstract}

%
%
%
%
%

\section{Introduction}
 The one-dimensional nonlinear Schr\"odinger equation (NLSE) is a nonlinear dispersive wave equation frequently used to describe wave propagation in optics and hydrodynamics~\cite{Agrawal,CHABCHOUB}. Besides, the NLSE also naturally arises in the study of several other physical systems, such as in the dynamics of the condensate wave function in BEC, as a model describing kinematics of vortices in liquid Helium, and  macromagnetic excitations in ferromagnets, to name a few~\cite{bec,hasimoto,lakshmanan}. Further, it is a completely integrable model with soliton solutions~\cite{shabat}. NLSE remains one of the well investigated models in the subject of solitons, and nonlinear dynamics in general, which also adds to   its pedagogical significance~\cite{faddeev}. Nevertheless, in spite of the rather elaborate literature on the subject, the NLSE continues to be a rich source for unanticipated phenomena. For instance, {\it rogue} wave behavior in NLSE has been a major subject of curiosity in its own right since theoretical results were first reported in 1983~\cite{peregrine}. Since then the phenomenon has been the subject matter of several investigations in diverse areas from water waves to optics, and is predicted to occur in BEC~\cite{kharif2008rogue,solli,bludov}. It is now understood that the {\it rogue} is a special case of the more general {\it breather} mode, witnessed either as a spatially or temporally periodic  localized excitation~\cite{kuznetsov,ma1979,Akhmediev}. The NLSE has also been studied for its of modulation instability (MI), where weak periodic perturbations on a continuous wave background undergo growth-decay cycles~\cite{benjamin,Zakharov}. This nonlinear process is closely related to Fermi-Pasta-Ulam (FPU) recurrence~\cite{lake,van}. MI can also be described analytically by various \textit{breather} solutions of 1-D NLSE, owing to its integrability~\cite{Akhmediev}. 

While the integrable limit has a significance of its own, actual experimental systems are dissipative. Such a dissipative system can exhibit nontrivial localized dynamical structures, similar to solitons of the integrable NLSE, when an external driving is added to compensate for the loss. These self-organized dynamical objects are commonly called {\it dissipative solitons}~\cite{ds}. They have been realized experimentally in optical fiber cavities~\cite{leo} and microresonators~\cite{kippenberg,herr2014}. Lately, the formation of dissipative Kerr solitons in optical microresonators were  identified as states in Kerr-frequency combs thus making them useful for practical applications~\cite{herr2014}. The underlying physical mechanism responsible for the formation of soliton pulses in these systems is  four-wave mixing. Mathematically, such a system can be described by the Lugiato-Lefever equation (LLE) which is a driven, damped  NLSE~\cite{LLE,barashenkov1996}. The equation was originally introduced to study spatially localized structures in  driven nonlinear optical systems~\cite{LLE}.

Another frequently studied model for self-organization phenomenon in nonlinear dissipative systems is the parametrically driven, damped NLSE. It has several applications. For instance, the equation models the parametric excitation of spin waves in ferromagnets and dynamics of small amplitude breathers in a long Josephson junction~\cite{gluzman1994,barashenkov1991}. Furthermore, under conditions when driving and dissipation are balanced, the equation can also exhibit solitary wave solutions~\cite{barashenkov1991,miles1984}. These parametric excitations are observed in water tanks when the oscillations are driven by periodically varying a parameter of the system at twice its frequency~\cite{wu1984,wang1997}. In optics, they come under the class of dissipative solitons and are known to occur in microresonators where the parametric driving originates from the second order term in nonlinear polarization~\cite{Longhi:95}. Although the existence of dissipative solitons in these systems require an external driving to counteract dissipation, there exist wide classes of traveling solitons for the undamped parametrically driven NLSE~\cite{barashenkov2001}.

In all the aforementioned physical situations, the solutions were either a stationary or a moving localized soliton. Even though breather solutions were obtained in the numerical simulation of LLE~\cite{yu2017}, the effect of driving on periodic breather solutions of NLSE and their stability have not been completely understood. In particular, the parametric driving of breather solutions remains largely unexplored. Hence,  in this work, we numerically study the evolution of a spatially periodic breaher solution of NLSE - the Akhmediev breather - under parametric driving. We show that, for certain range of parameter values, the initial breather profile travels with a constant speed without decreasing its amplitude, curiously like a soliton. Moreover, we observe that the dynamics show noticeable differences with a 1-soliton of NLSE when dissipation is included. Finally, we also discuss the stability of these solutions under a variety of random perturbations.

\section{Dynamics of small-amplitude Akhmediev breathers under parametric driving}

\begin{figure}
\centering
\begin{framed}
\begin{subfigure}{0.49\textwidth}
\centering
\includegraphics[width = 1 \textwidth]{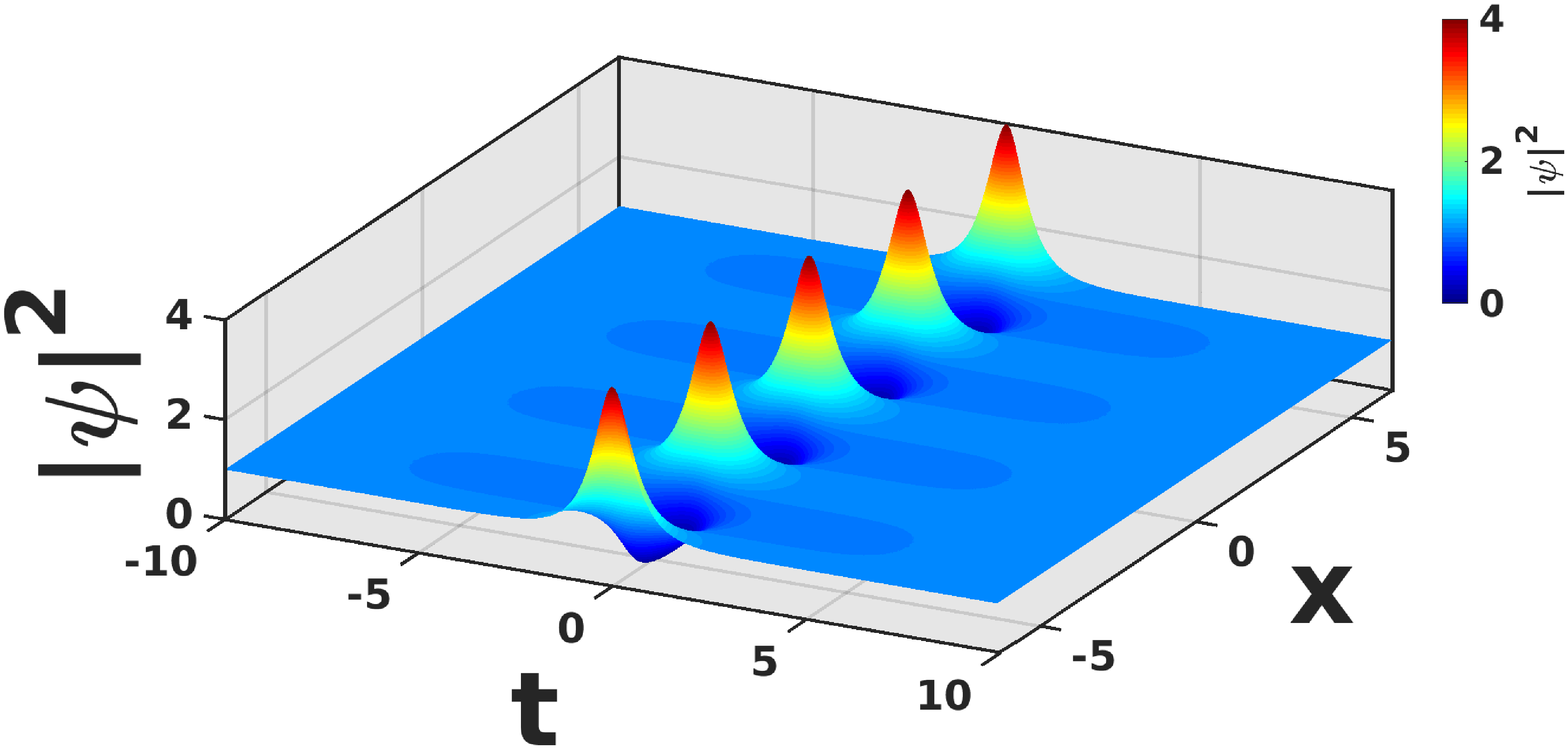}
\caption{}
\label{figure1a}
\end{subfigure}
\begin{subfigure}{0.49\textwidth}
\centering
\includegraphics[width =1\textwidth]{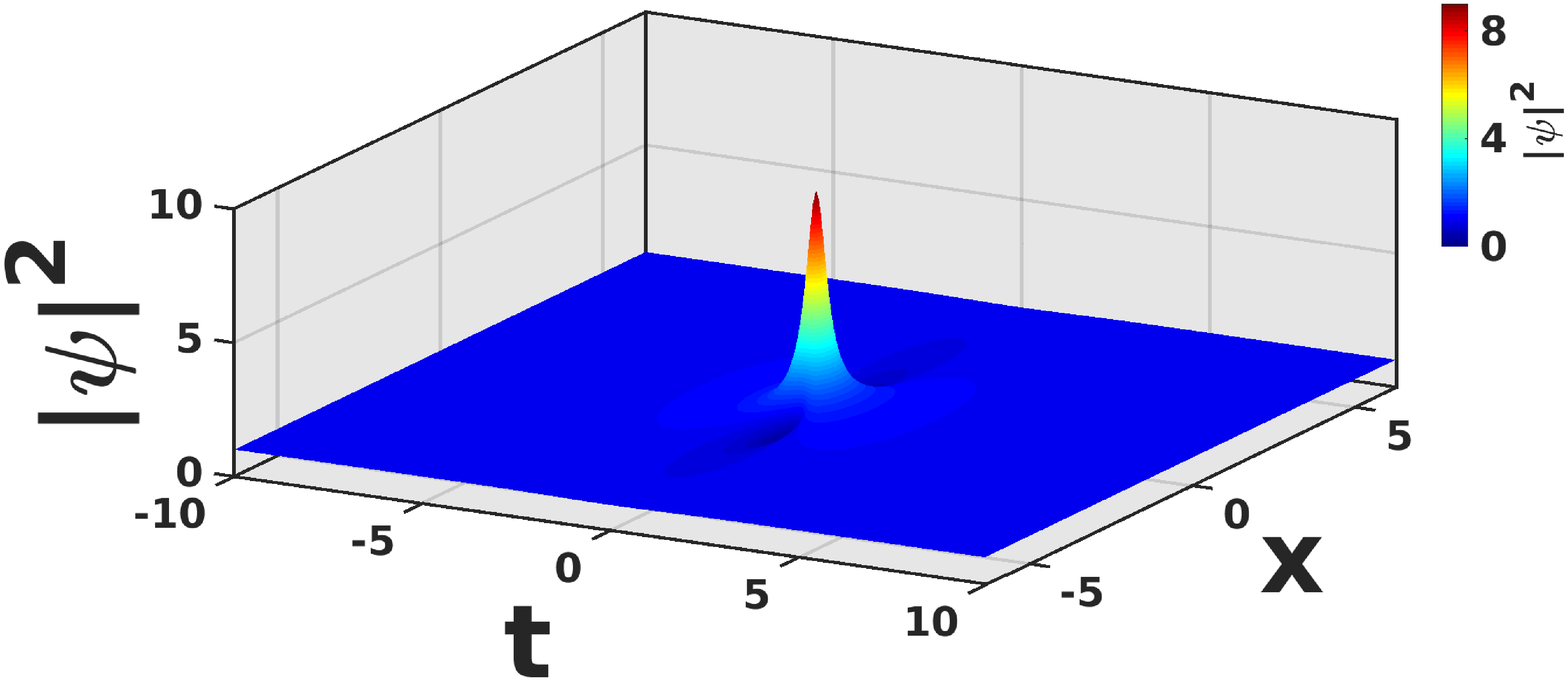}
\caption{}
\label{figure1b}
\end{subfigure}
\caption{(a) Akhmediev breather for the parameter $\xi=0.5$. (b) Peregrine {\it `rogue'} soliton }
\label{figure1}
\end{framed}
\end{figure}
In dimensionless form, the 1-D NLSE is given by
\begin{equation}
\label{eq1}
\centering
i\psi_t+\psi_{xx}+2|\psi|^2\psi=0
\end{equation} 
where $\psi(x,t)$ is a complex field and the variables $x$ and $t$ refer to dimensionless \textit{space} and \textit{time} respectively. In the context of propagation of light through optical fibers, $\psi(x,t)$ is the complex amplitude of electric field and the variables $t$ and $x$ correspond to the propagation distance and time respectively. Equation (\ref{eq1}) when supplemented with initial and boundary conditions can be solved for soliton solutions by any of the standard methods, such as the Darboux transformation. For example, the first order periodic solution

\begin{equation}
\label{eq2}
\eqalign{\psi_{AB}(x,t)=\e^{i2t} \left(1+\xi\frac{2\cos(qx)-2\xi\cosh(\Omega t)+iq\sinh(\Omega t)}{\cosh(\Omega t)-\xi \cos(qx)}\right), \cr \Omega=q \sqrt{4-q^2},\hspace{0.5cm} q=2\sqrt{1-\xi^2}} 
\end{equation}
could be generated from an initial {\it seed} plane wave solution, $\e^{i2t}$. For $0<\xi<1$, equation (\ref{eq2}) is the Akhmediev breather (AB) which is a train of localized pulses in the intensity profile, $|\psi|^2$, that are located periodically along the $x-$axis with period $\frac{\pi}{\sqrt{1-\xi^2}}$~(figure~(\ref{figure1a})). The value of the parameter $\xi$ determines the degree of localization. When $\xi=0$, $\psi_{AB}$ reduces to the continuous wave (cw) solution, $\e^{i2t}$, and the limit $\xi\rightarrow1$ produces the Peregrine soliton~(figure~(\ref{figure1b})) - a localized solution in both space ($x$) and time ($t$). The AB can be viewed as the result of an instability of the cw solution, thus providing an analytic description for MI.


 In this paper, we discuss the evolution of AB in the parametrically driven NLSE
 
\begin{equation}
\label{eq3}
\centering
i\psi_t+\psi_{xx}+2|\psi|^2\psi=f(x,t)\psi^*-i\beta \psi
\end{equation}
where $f(x,t)=f_0 \exp(iKx)$ is a driving force, $f_0$ and $K$ are constants, and $\beta>0$ is the dissipation. It may be noted that the transformation
\begin{equation}
\psi(x,t) = \Psi(X,t) \ \e^{iKx/2}
\end{equation}
\label{eq4}
to a moving frame $X = x - Kt$ leads to the equation
\begin{equation}
\label{eq5}
i\Psi_t+\Psi_{XX}+2|\Psi|^2\Psi=f_0\Psi^* -i\beta \Psi+ \frac{K^2}{4} \Psi
\end{equation}
with an additional detuning term. Equation (\ref{eq5}) naturally arises in an optical microresonator containing a Kerr medium, where the parametric driving is realised using a  nonlinear $\chi^{(2)}$ medium~\cite{Longhi:95}. 

\subsection{The nondissipative case}
\begin{figure}
\centering
\begin{framed}
\begin{subfigure}{0.49\textwidth}
\centering
\includegraphics[width = 1 \textwidth]{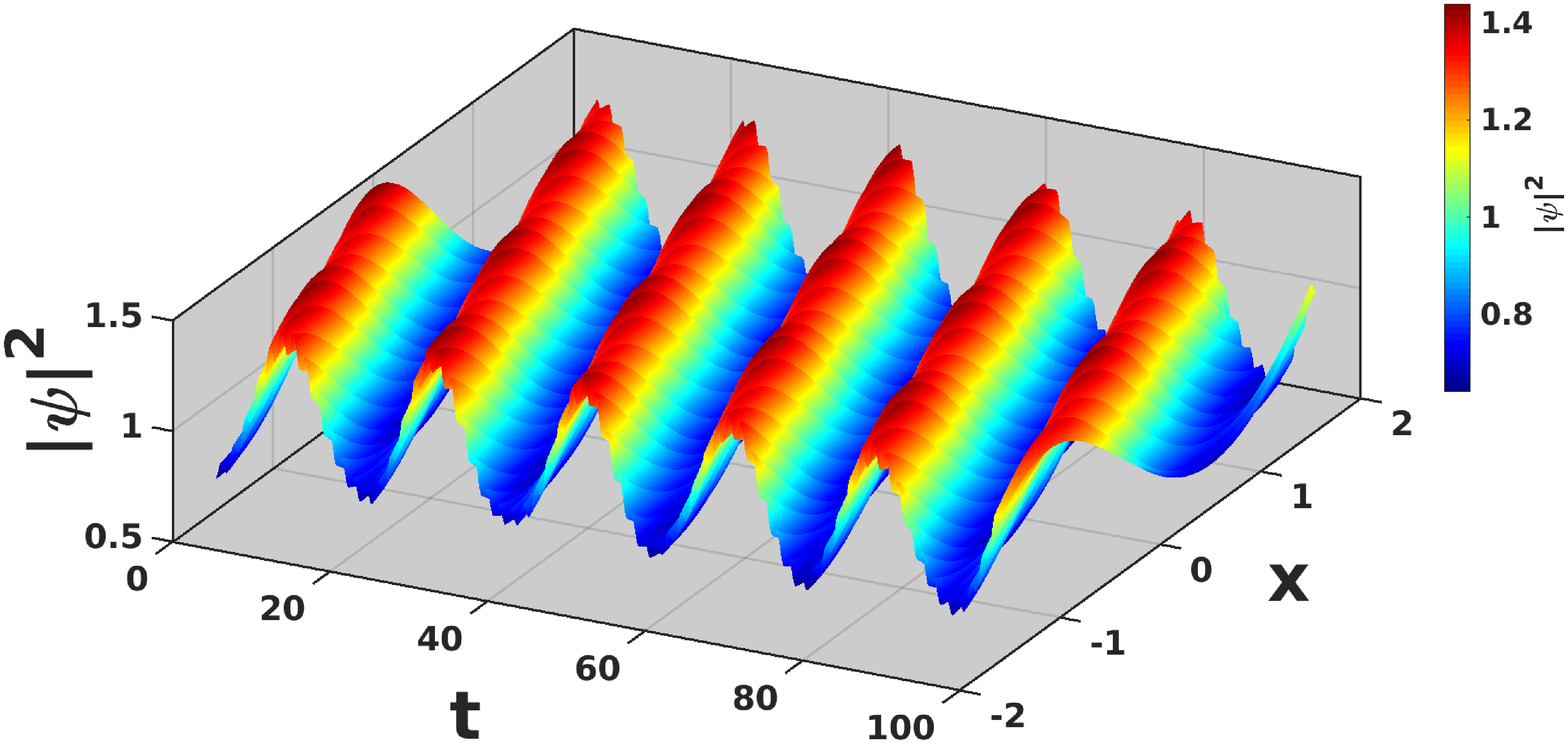}
\caption{}
\label{figure2a}
\end{subfigure}
\begin{subfigure}{0.49\textwidth}
\centering
\includegraphics[width = 1 \textwidth]{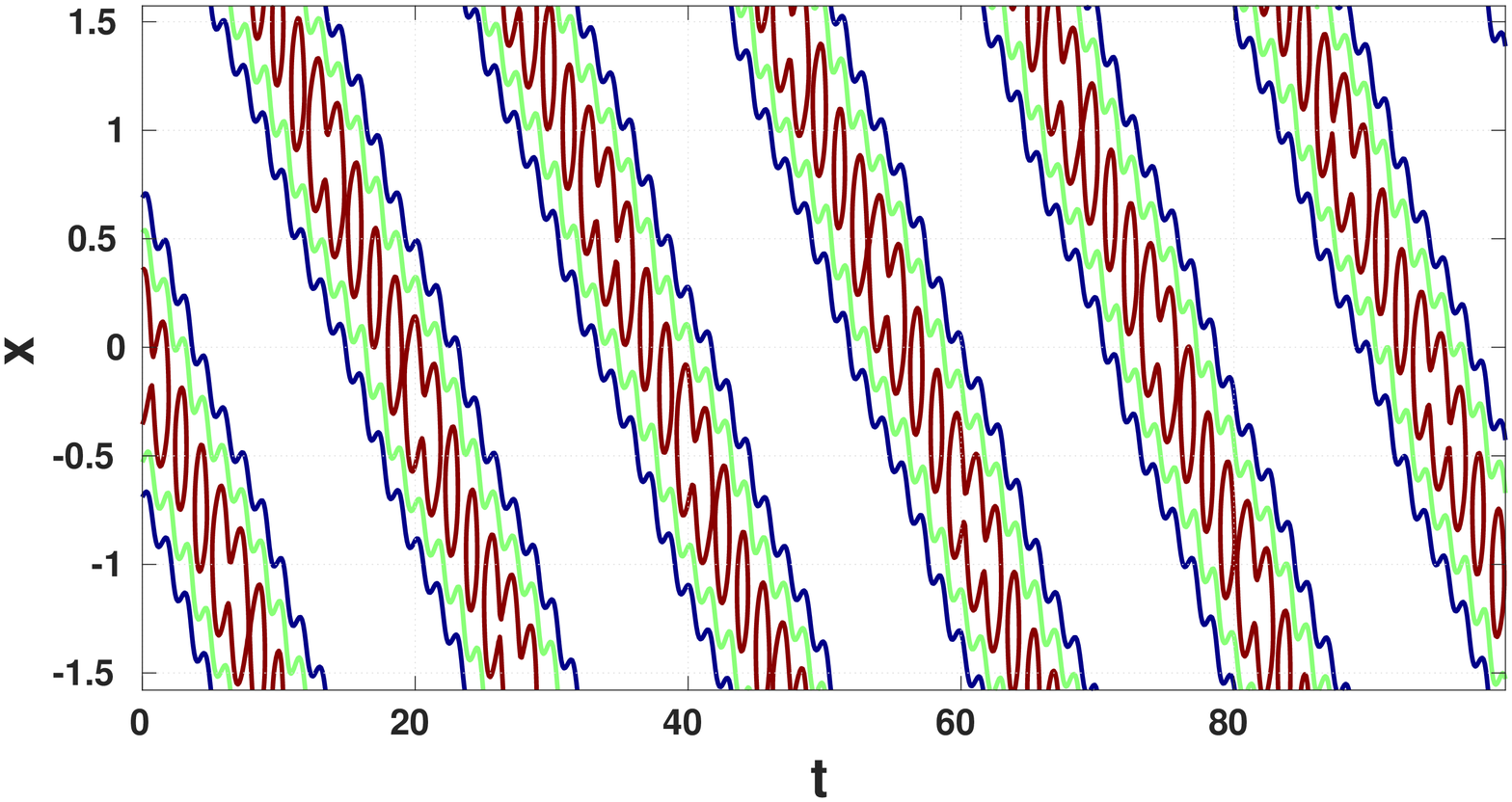}
\caption{}
\label{figure2b}
\end{subfigure}
\begin{subfigure}{0.49\textwidth}
\centering
\includegraphics[width =1\textwidth]{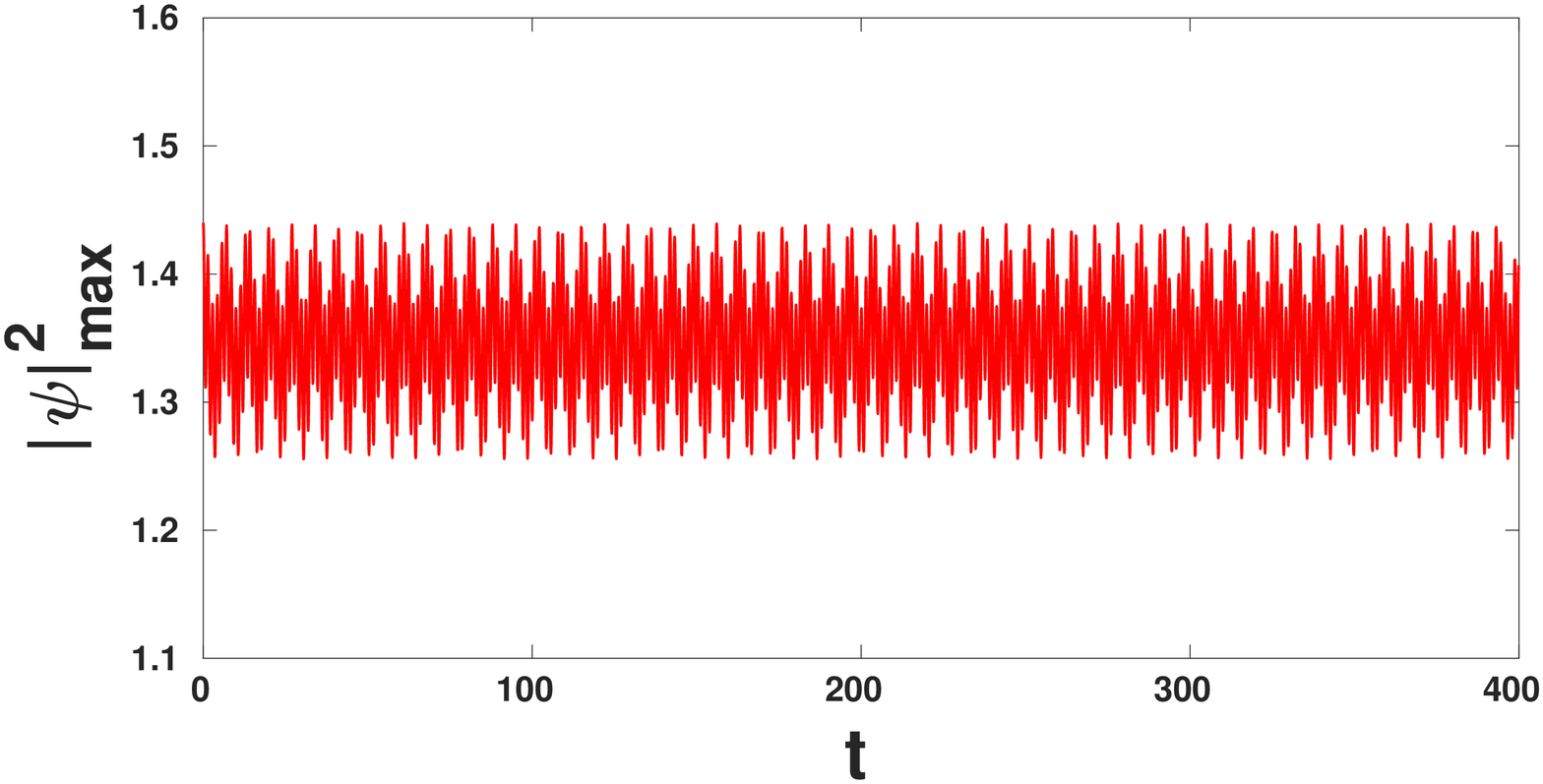}
\caption{}
\label{figure2c}
\end{subfigure}
\begin{subfigure}{0.49\textwidth}
\centering
\includegraphics[width = 1 \textwidth]{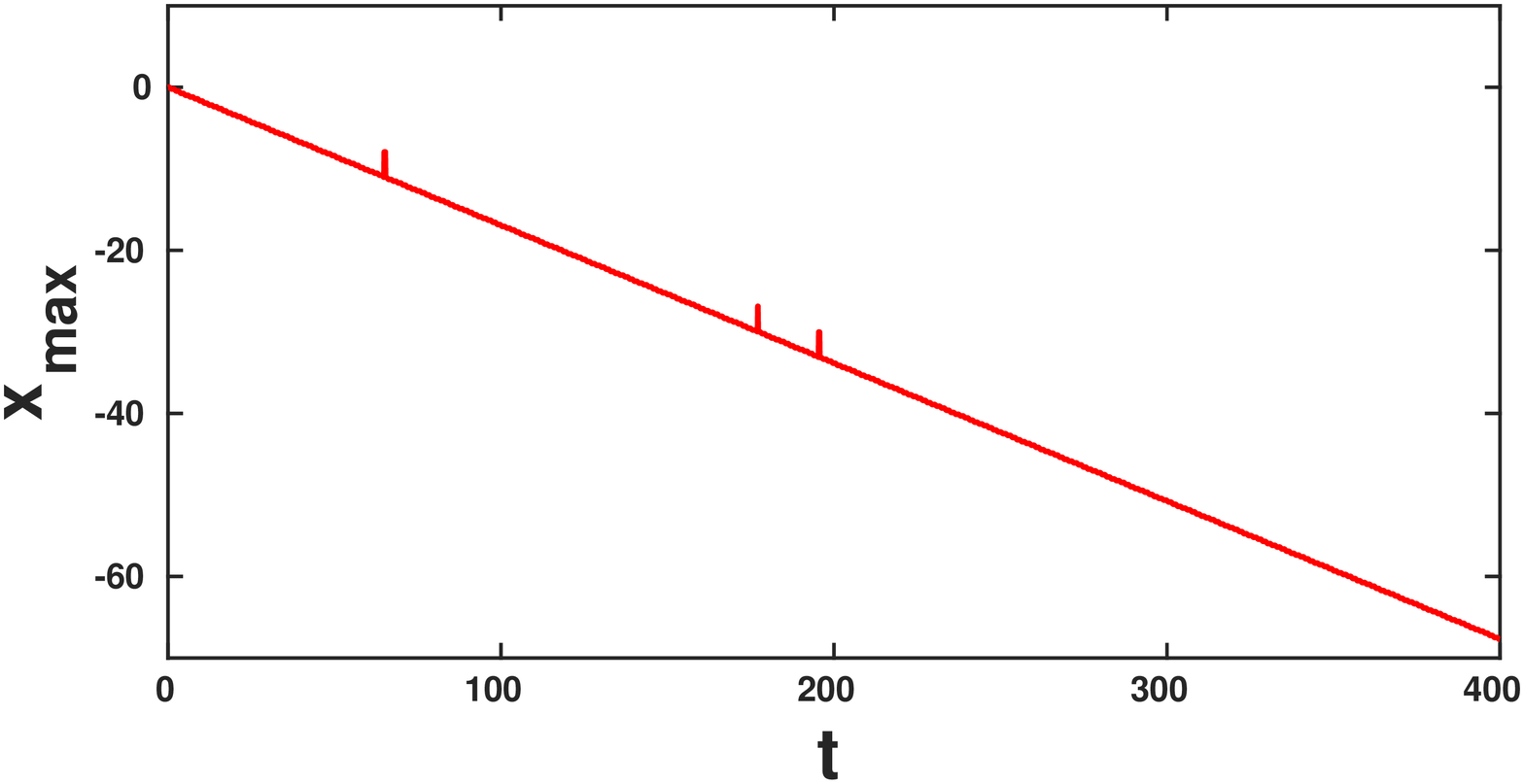}
\caption{}
\label{figure2d}
\end{subfigure}
\caption{(a) Soliton-like behaviour observed in the evolution of AB under parametric driving and zero dissipation. (b) Contour plot of (a). (c) Evolution of the intensity maximum. (d) Position of the intensity maxima vs time. The forcing has a magnitude $f_0=0.05$. The external driving has the same periodicity as that of the lattice. Parameters used are $\kappa=1$ and $\xi=0.1$. }
\label{figure2}
\end{framed}
\end{figure}
\begin{figure}
\begin{framed}
\centering
\includegraphics[scale=0.3]{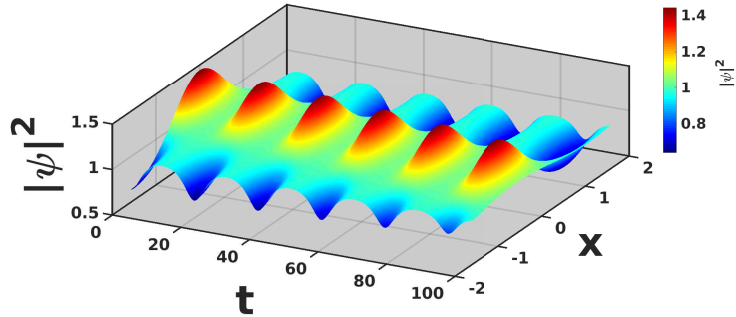}
\caption{Intensity plot when $K=10$ and $q=1.99$ (or $\xi=0.1$). Here, unlike in the $K=q$ case, the dynamics is characterized by the appearance of maxima that locally resemble breather excitations. The other parameters are as in figure (\ref{figure2}).}
\label{figure3}
\end{framed}
\end{figure}

\begin{figure}
\begin{framed}
\centering
\includegraphics[scale=0.39]{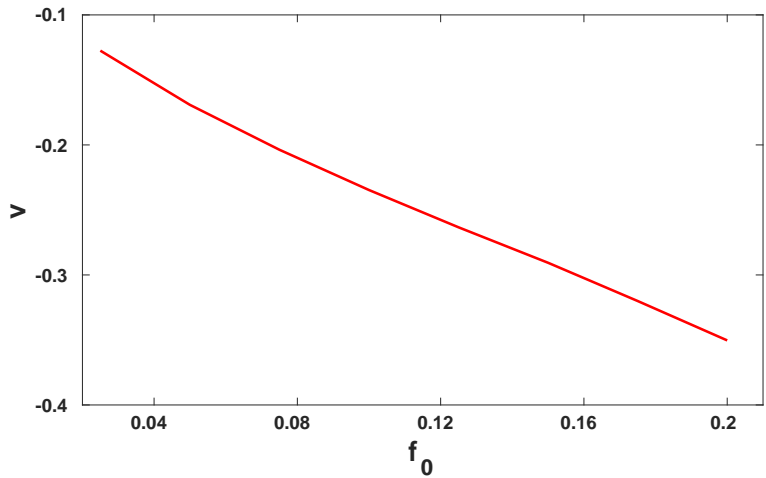}
\caption{Velocity vs $f_0$ for $\xi=0.1$. We observe that the speed of the solution increases with $f_0$ until a threshold value of $f_{0,max}=0.2$. The lower limit of $f_0$, to observe soliton solution, for this particular initial condition  is found to be $0.025$. }
\label{figure4}
\end{framed}
\end{figure}

\begin{figure}
\begin{framed}
\centering
\includegraphics[scale=0.39]{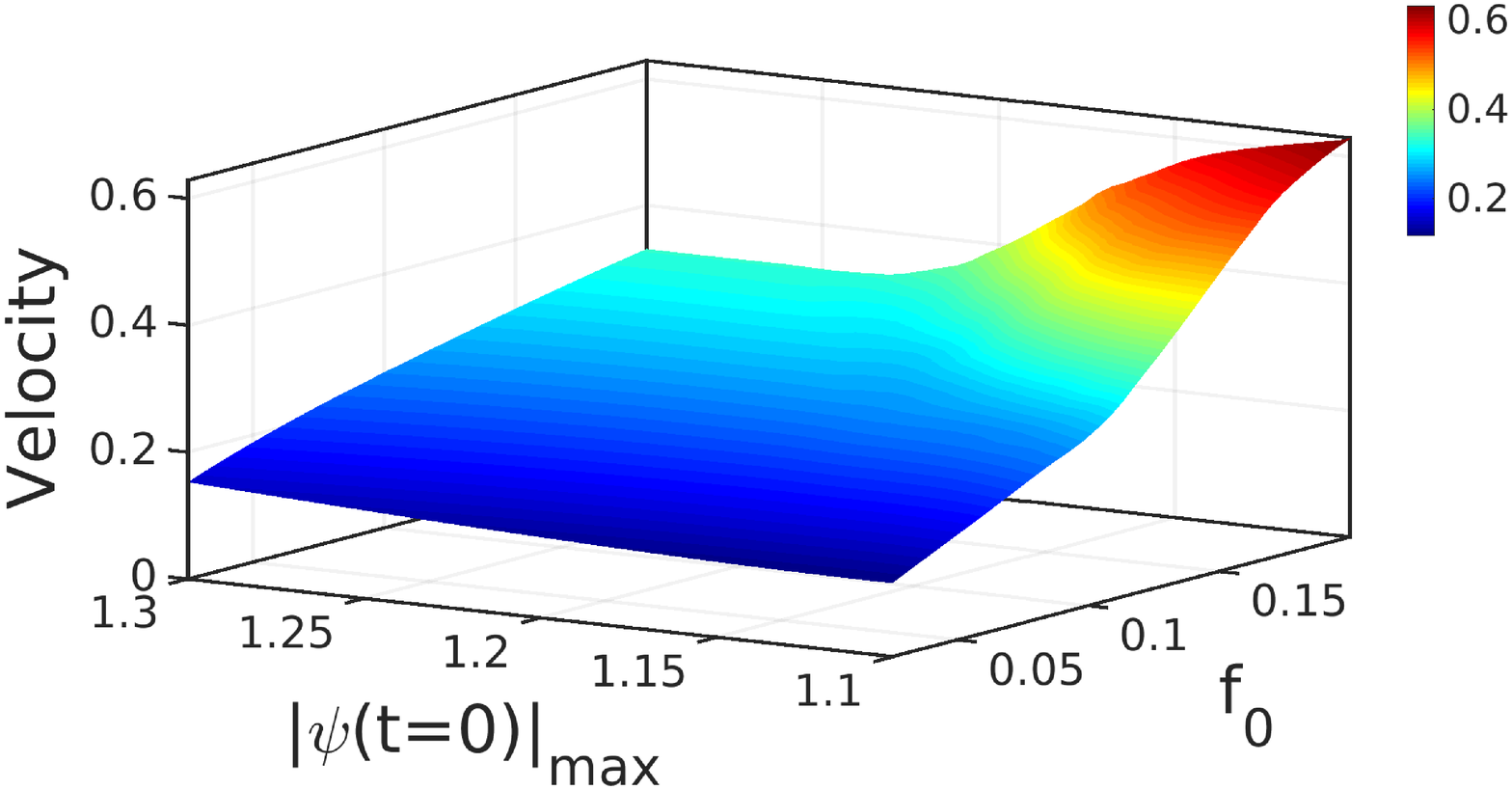}
\caption{Velocity as a function of maximum of initial amplitude and magnitude of forcing. In the blue region where the velocity is small, the evolution is soliton-like. Here, the initial amplitude retains its shape, making the initial waveform a very good approximation to the soliton profile at later time $t$. For large values of $f_0$, we see that the velocity decreases with increase in the initial amplitude (red region). This apparent anomaly in the velocity is due to the onset of a breather-like behaviour.}
\label{figure5}
\end{framed}
\end{figure}

In this section, we describe the evolution of AB in the parametrically driven NLSE with zero damping, i.e., $\beta=0$. To this end, we numerically intergate (\ref{eq3}) using the initial condition 
\begin{equation}
\label{eq6}
\psi_0\equiv\psi_{AB}(x,0)=\left(1+\xi\frac{2\cos(qx)-2\xi}{1-\xi \cos(qx)}\right)
\end{equation}
and periodic boundary condition
\begin{equation}
\label{eq7}
\psi(x+L,t)=\psi(x,t).
\end{equation}

The numerical simulations are performed using the split-step Fourier method (SSFM)~\cite{weideman}. Here, the nonlinear equation is split into two parts 
\begin{eqnarray}
\psi_t= i\psi_{xx} \label{eq8} \\
\psi_t = i(2|\psi|^2\psi-f(x,t)\psi^*) \label{eq9}
\end{eqnarray} 
wherein, the solution is advanced from $t$ to $t+\delta t$ in two steps. In the first step, we solve (\ref{eq8}) in the Fourier domain with the initial condition $\psi(x,t)$. The resulting solution is used as the initial value to solve (\ref{eq9}), yielding the final solution, $\psi(x,t+\delta t)$, at $t+\delta t$. In the second step, we employ the fourth order Runge-Kutta method. All simulations are performed using a spatial lattice consisting of $512$ points and a time step $\delta t= 10^{-4}$.

As described in the beginning of this section, the Akhmediev breather is a periodic and localized excitation. It is naturally expected that this behavior prevails in the driven damped case, for a suitable choice of the driving force
and damping balancing each other. In the absence of the  balancing damping term,  the external forcing is expected to render the breather mode unstable.
 Yet interestingly, we observe a new localized waveform that travels like a usual 1-soliton solution. This is in contrast to the unperturbed breather dynamics where localizations eventually decay to the continuous wave background. Figure~(\ref{figure2a}) shows the intensity plot obtained by choosing the initial condition (\ref{eq6}). Here, the initial breather profile travels with a constant velocity without much reduction in its amplitude. The velocity of the solution can be obtained by plotting the position of maxima versus time. In figure~(\ref{figure2d}), the locations of maxima are plotted against time for the parameter $\xi=0.1$ and $f_0=0.05$. We define the average velocity of the solution as the slope of this line. It is noted that this soliton-like behavior is only observed when the frequency of the driving force is equal to the wavenumber of the initial breather profile ($K=q$). The case $K\neq q$ is described in figure(\ref{figure3}). A few remarks about the new soliton structure are in order. Firstly, the dynamics depicted in figure (\ref{figure2a}) shows that the initial waveform retains its shape during evolution. Thus  the resemblance to a 1-soliton solution is only qualitative and cannot be regarded as a breather to soliton conversion as in~\cite{mahnke2012}. This is because a soliton with energy of the order of the breather is a more localized solution with higher amplitude. In other words, the ratio of peak to full width at half maximum (FWHM) for a 1-soliton is higher compared to a breather of the same energy.

Second, the occurrence of soliton-like solution is restricted to small amplitude breathers which is determined by the parameter $\xi$. Figure (\ref{figure2}) shows the results of numerical simulation when the parameter $\xi=0.1$. As we increase the amplitude of the initial breather, a large number of peaks with varying intensity appears in the intensity profile at random locations. Our numerical simulations could not confirm any stable pattern in the long time dynamics of higher amplitude ABs.   

Finally, the magnitude of the forcing also plays an important role in the dynamics. For small values of $\xi$, the initial breather profile travels with a constant velocity when the magnitude of forcing $f_0$ is in a certain range (determined numerically). The velocity can be positive or negative depending  on the sign of $f_0$ ; when $f_0$ is positive (negative) the breather travels in the negative (positive) $x$-direction. As we increase the value of $f_0$, the speed of the solution increases until a threshold $f_{0,max}$ is reached beyond which the pattern is destroyed. Figure (\ref{figure4}) shows the variation of velocity with the magnitude of forcing for the initial breather defined by the parameter $\xi=0.1$. We also mention, in figure (\ref{figure5}), the region in $(f_0,|\psi(t=0)|_{max})$ plane where a soliton-like solution is observed.

\subsection{Effect of dissipation on the parametrically driven Akhmediev breather (PDAB)}
\begin{figure}
\centering
\begin{framed}
\begin{subfigure}{0.49\textwidth}
\centering
\includegraphics[width = 1 \textwidth]{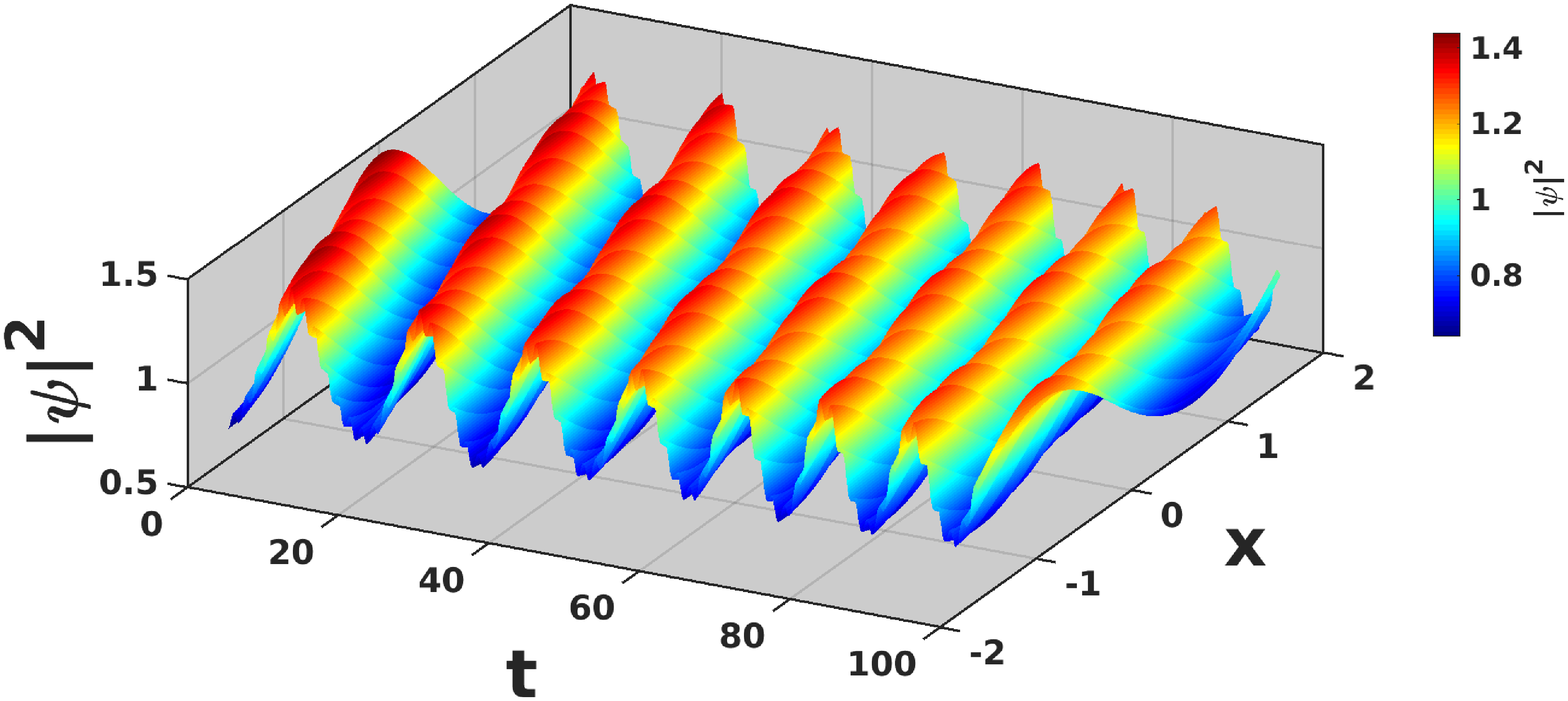}
\caption{}
\label{figure6a}
\end{subfigure}
\begin{subfigure}{0.49\textwidth}
\centering
\includegraphics[width = 1 \textwidth]{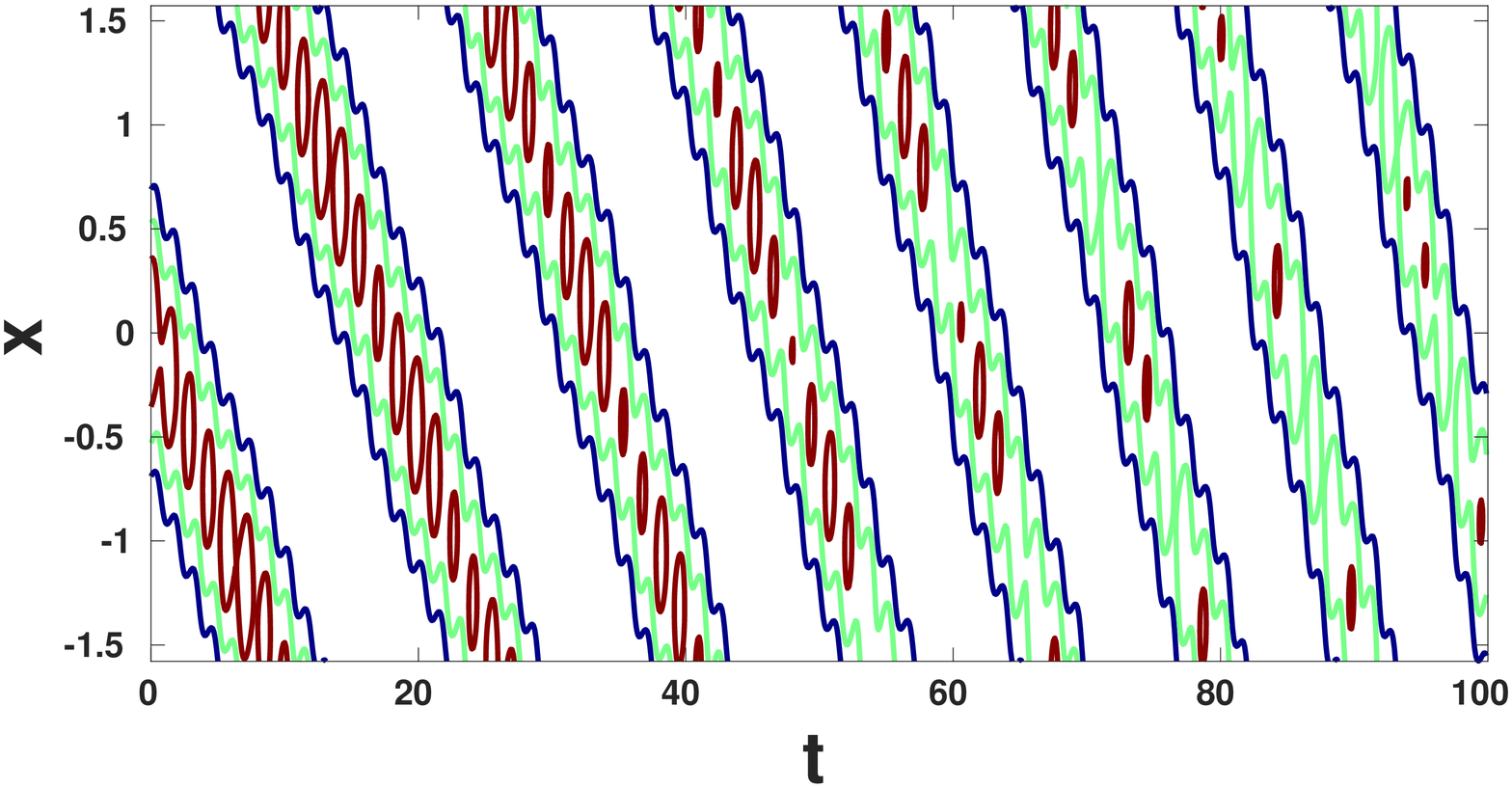}
\caption{}
\label{figure6b}
\end{subfigure}
\begin{subfigure}{0.49\textwidth}
\centering
\includegraphics[width =1\textwidth]{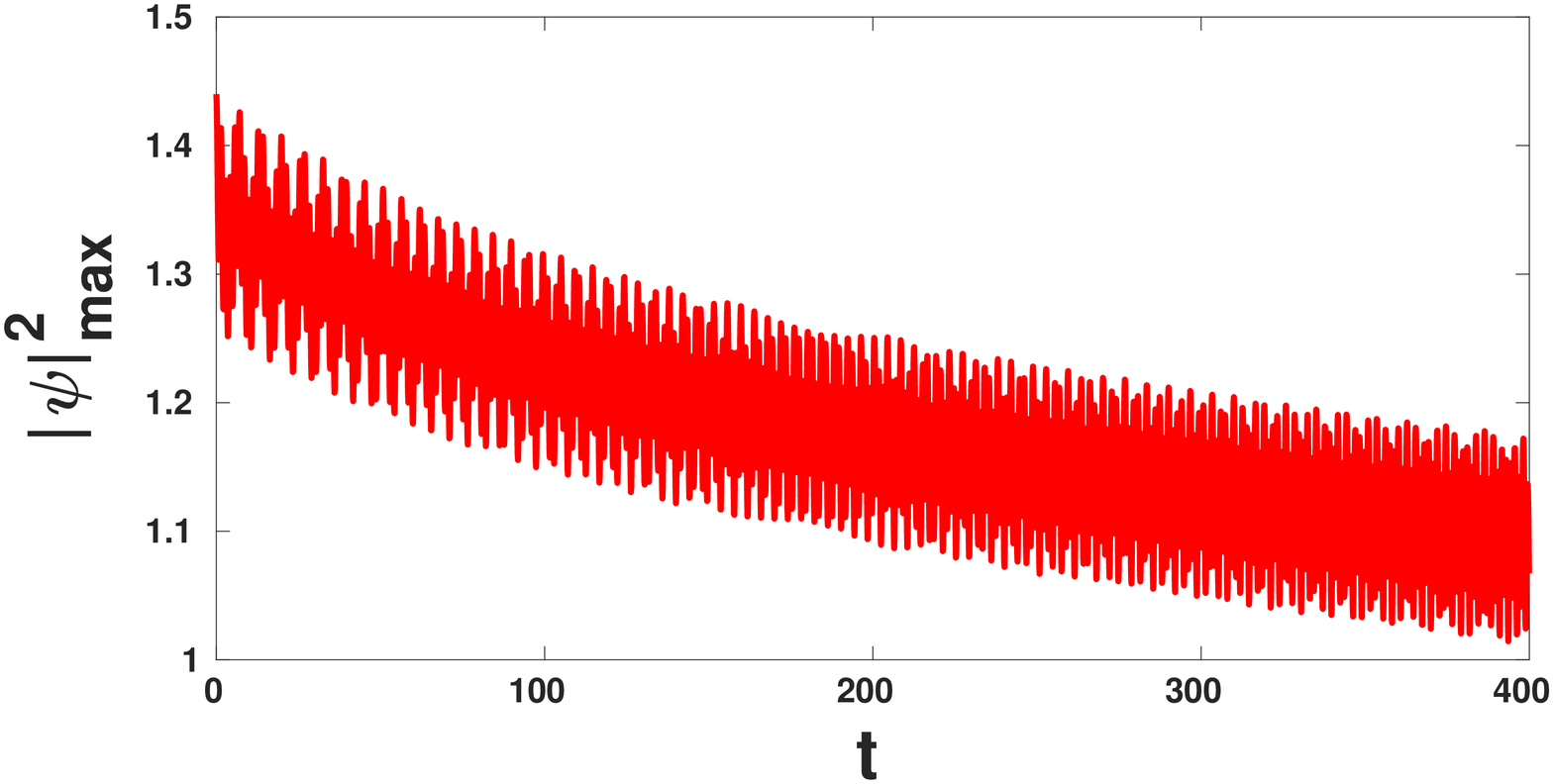}
\caption{}
\label{figure6c}
\end{subfigure}
\begin{subfigure}{0.49\textwidth}
\centering
\includegraphics[width = 1 \textwidth]{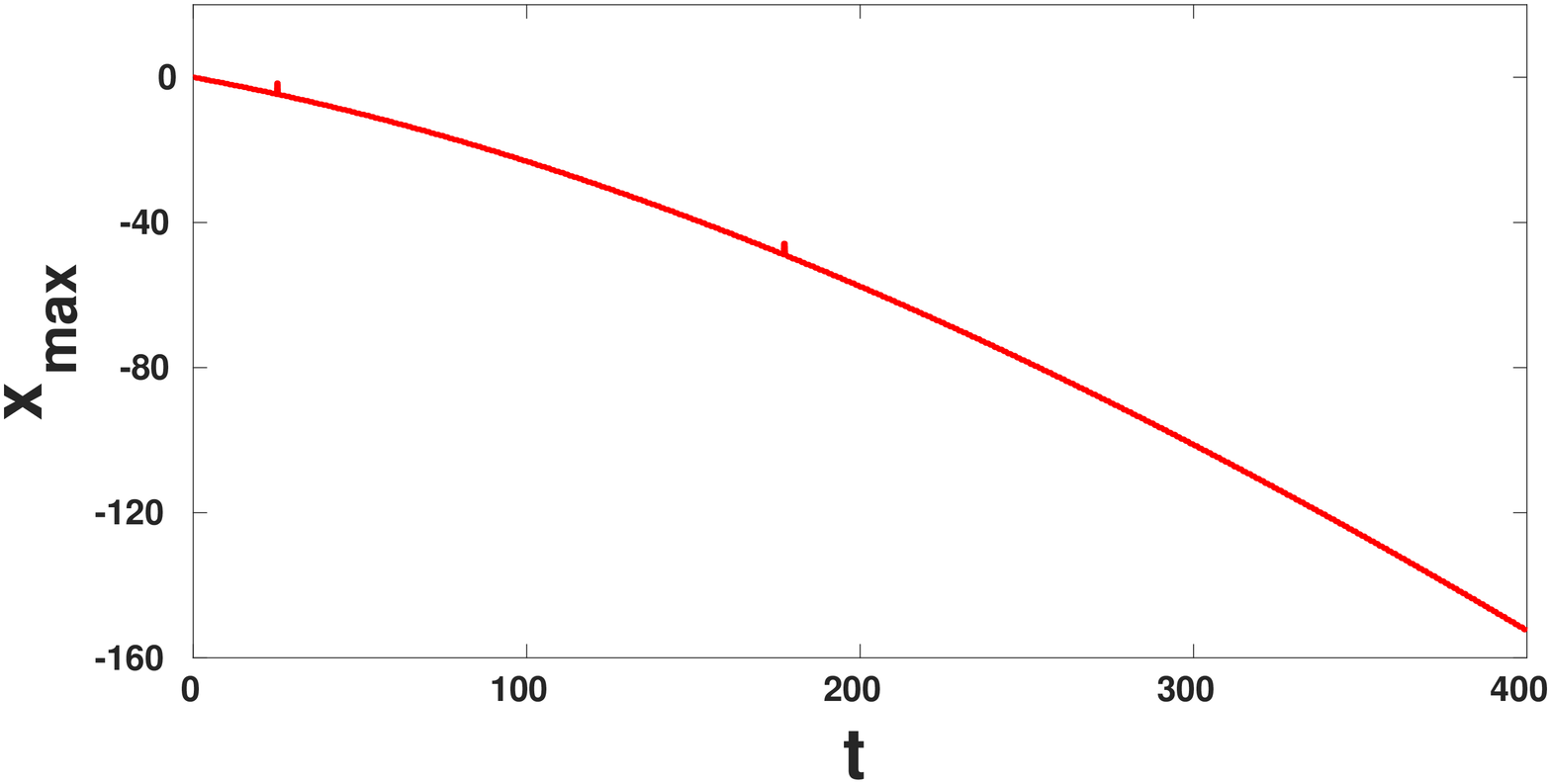}
\caption{}
\label{figure6d}
\end{subfigure}
\caption{(a) Effect of dissipation on the soliton-like solutions of figure(\ref{figure2}). (b) Contour plot of (a).  (c) Evolution of the intensity maximum. (d) Position of the intensity maxima vs time. Here, $\beta=10^{-4}$. The other parameters are as in figure (\ref{figure2}).}
\label{figure6}
\end{framed}
\end{figure}

For our model to apply to real physical situations, we must  invariably take into account dissipation. There are two different ways dissipation affects the solutions of a dynamical system:
\begin{enumerate}
\item It can cause stable solutions to exist such that the solution disappears when dissipation is turned off. An example is the dissipative solitons of driven, damped NLSE. Here, dissipation plays a principal role as the formation of dissipative solitons crucially depend on the balance between dissipation and driving aside from a balance of nonlinearity and dispersion. \item It can cause energy losses to an otherwise stable solution. For instance, when dissipation is included in NLSE, the amplitude of a soliton solution decreases as it propagates. Since the amplitude and velocity of solitons are linearly related, we can also observe a decrease in the velocity of the soliton. 
\end{enumerate}
In this section, we indicate the effects of dissipation on the paramterically driven Akhmediev breather (PDAB). We first note that, the PDAB is a stable solution in the absence of dissipation as confirmed in our numerical experiments. Hence we expect the solution to behave similar to the conservative NLSE soliton when dissipation is present. In fact, the amplitude of the PDAB solution decreases over time as shown in figure (\ref{figure6c}). However, in contrast to the conservative soliton, the velocity of the PDAB solution increases (fig (\ref{figure6d})). We remark that such a solution wherein an increase in velocity  along with a decrease in amplitude is counterintuitive and has not been observed before. The observation also brings about the following distinction; for a conventional conservative soliton, amplitude and velocity are linearly related, whereas the new soliton structure has an inverse relation between the two, for a range of parameter values. This intriguing characteristic emphasizes that the newly observed soliton-like solution is fundamentally different from the conservative soliton.

\section{Stability of the parametrically driven Akhmediev breather}

\begin{figure}
\centering
\begin{framed}
\begin{subfigure}{0.49\textwidth}
\centering
\includegraphics[width = 1 \textwidth]{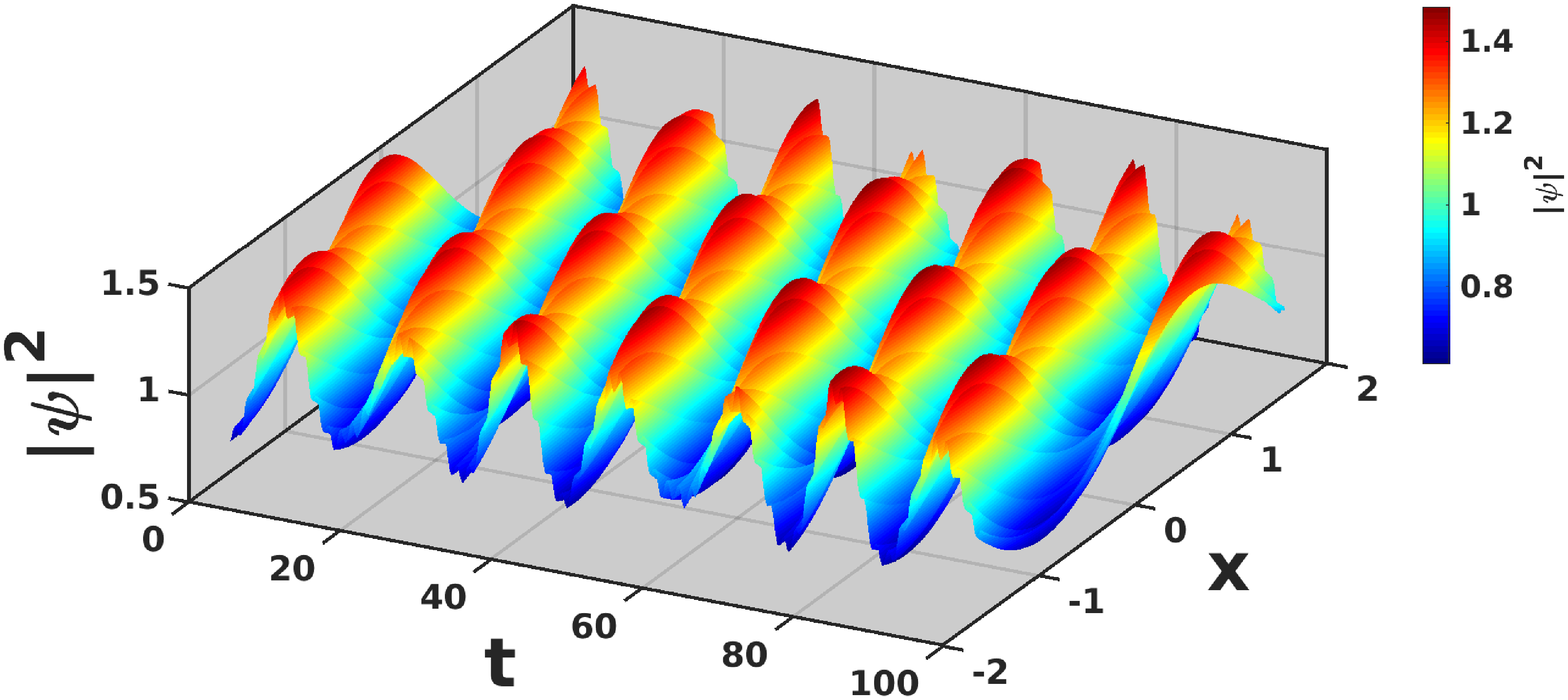}
\caption{}
\label{figure7a}
\end{subfigure}
\begin{subfigure}{0.49\textwidth}
\centering
\includegraphics[width = 1 \textwidth]{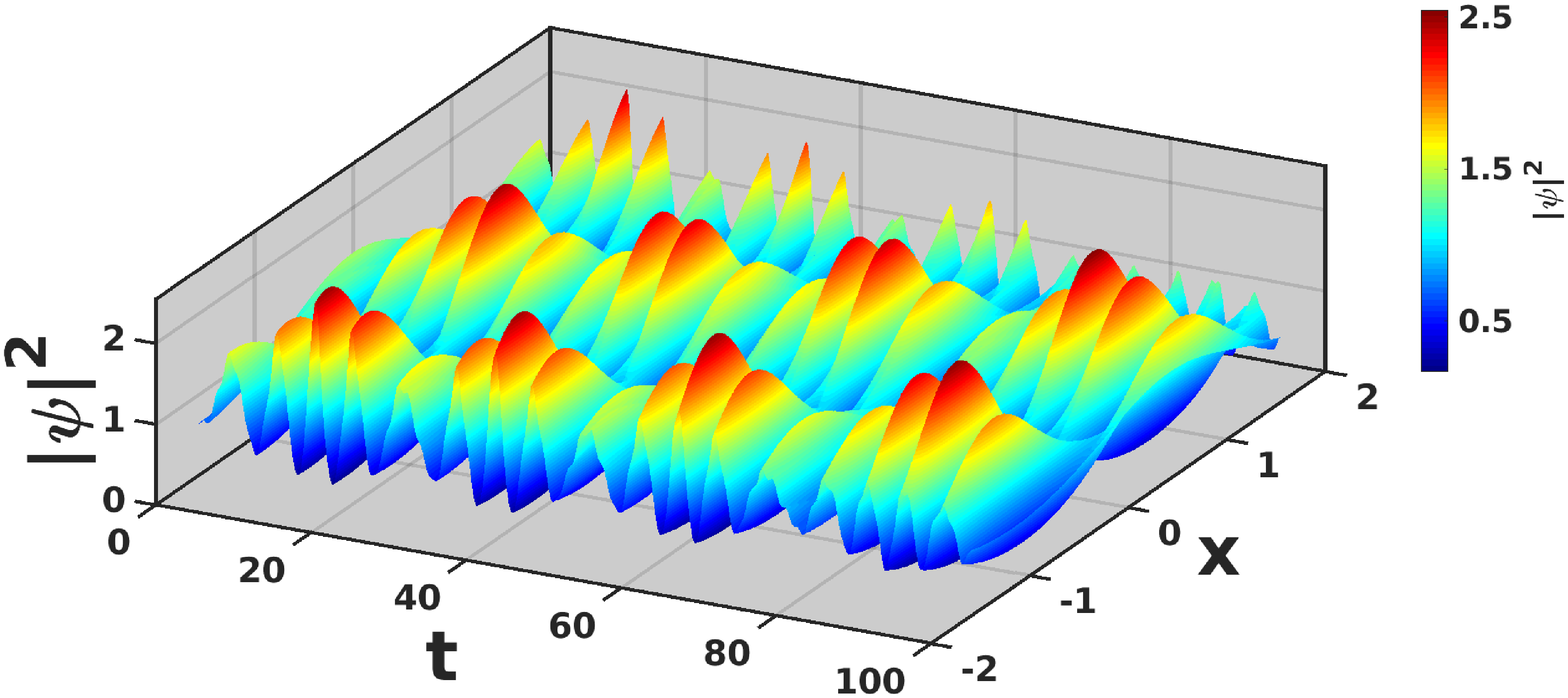}
\caption{}
\label{figure7b}
\end{subfigure}
\begin{subfigure}{0.49\textwidth}
\centering
\includegraphics[width = 1 \textwidth]{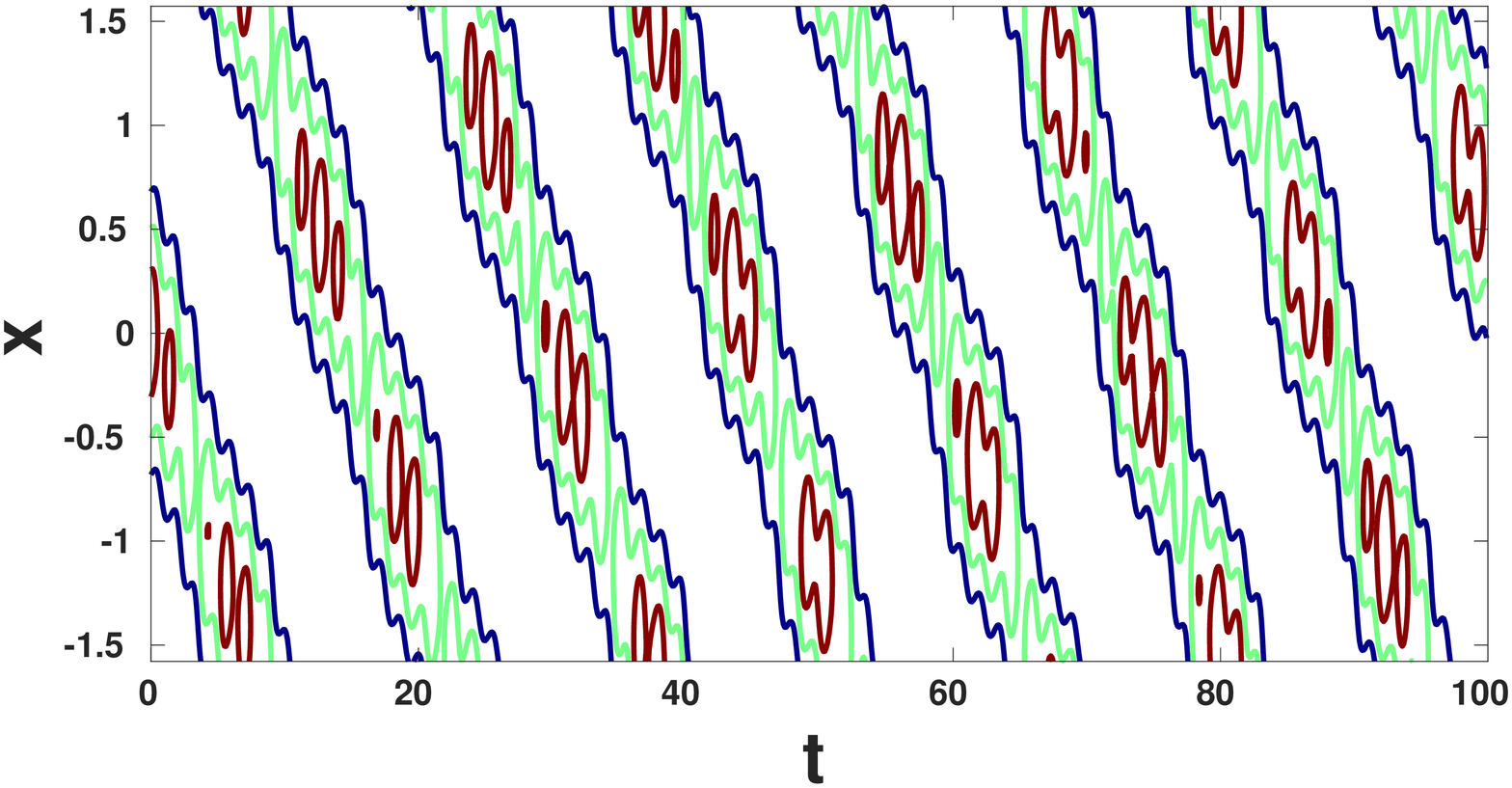}
\caption{}
\label{figure7c}
\end{subfigure}
\begin{subfigure}{0.49\textwidth}
\centering
\includegraphics[width = 1 \textwidth]{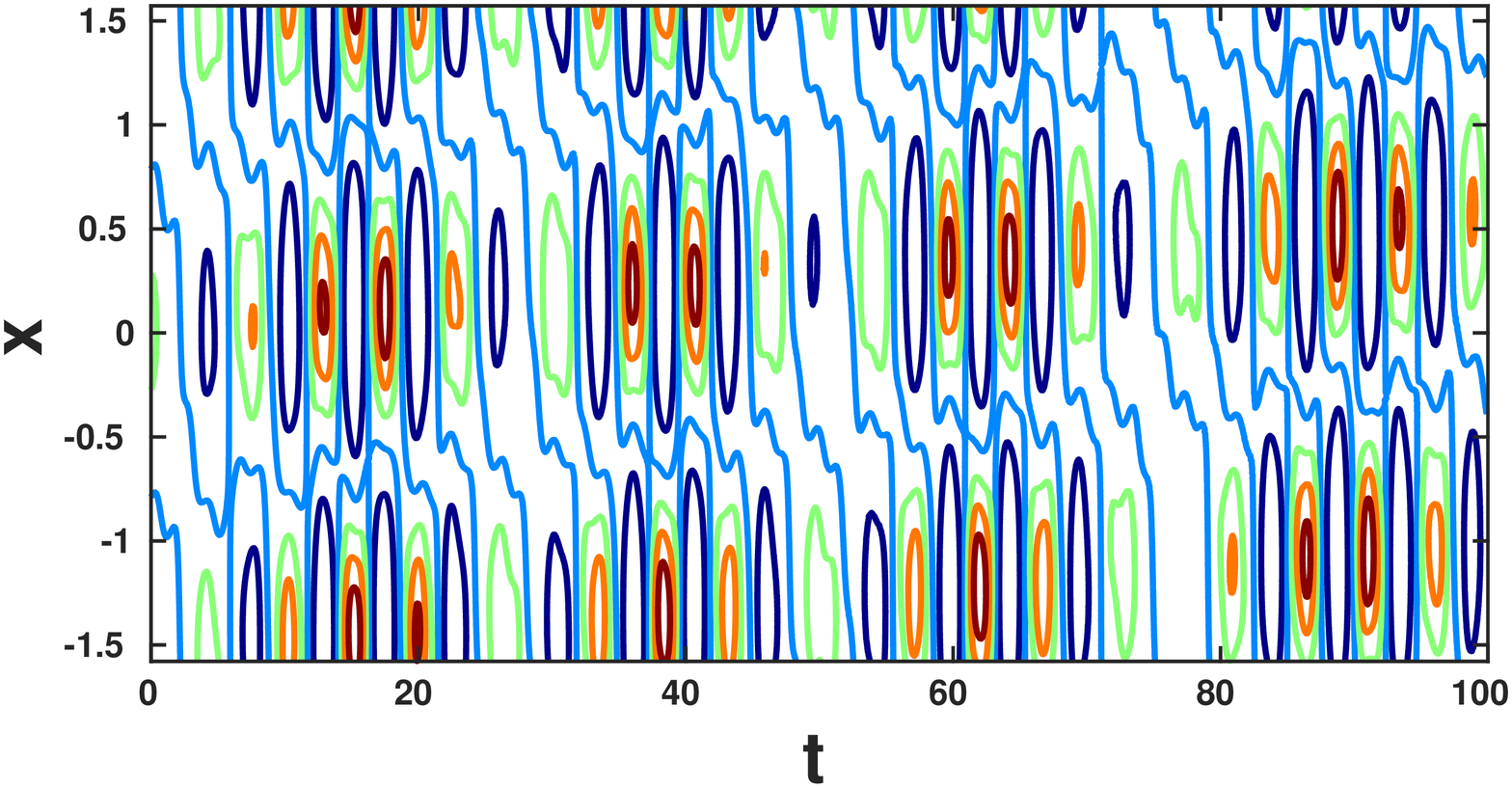}
\caption{}
\label{figure7d}
\end{subfigure}
\begin{subfigure}{0.49\textwidth}
\centering
\includegraphics[width =1\textwidth]{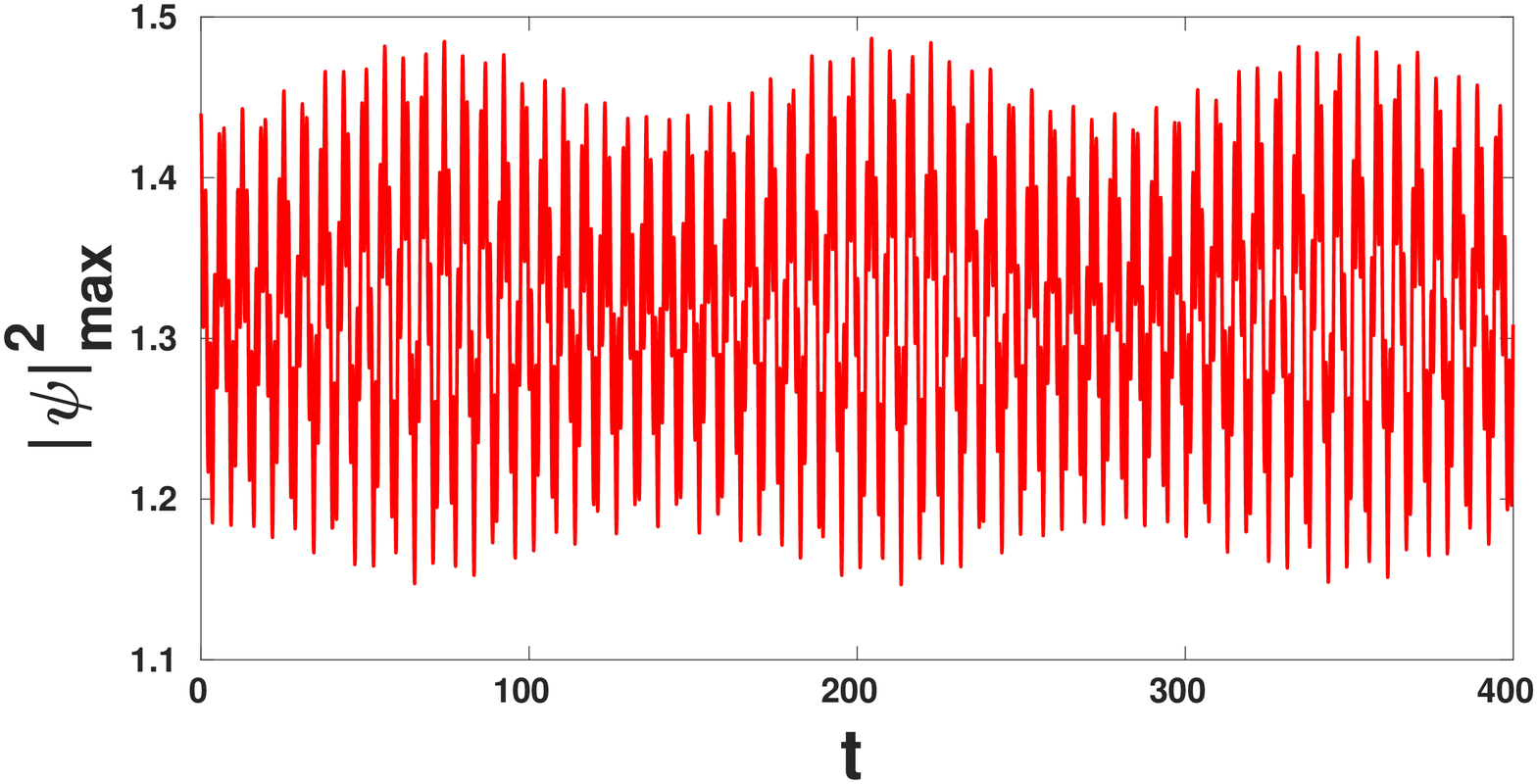}
\caption{}
\label{figure7e}
\end{subfigure}
\begin{subfigure}{0.49\textwidth}
\centering
\includegraphics[width =1\textwidth]{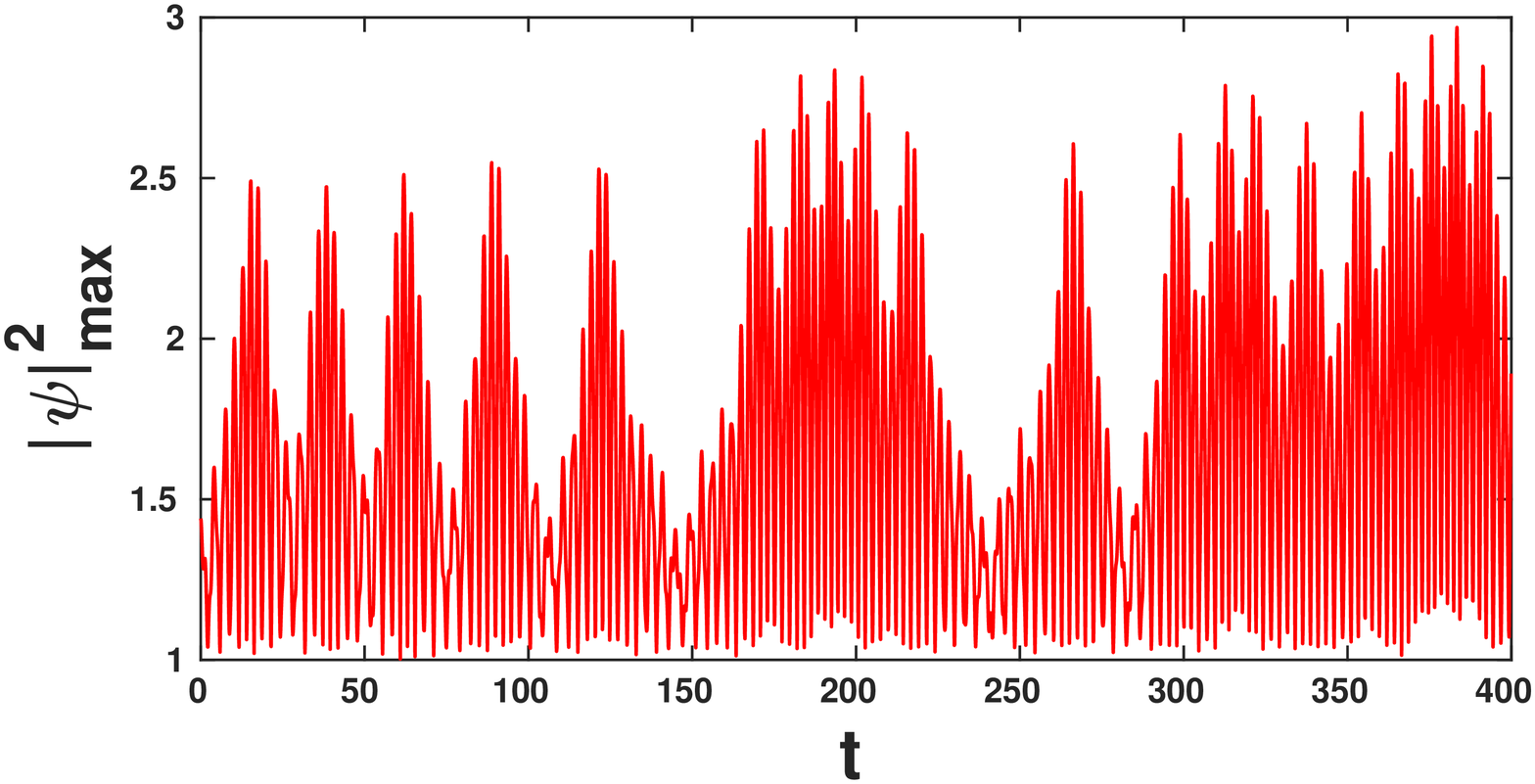}
\caption{}
\label{figure7f}
\end{subfigure}
\caption{Effect of additive white noise on the PDAB solution. (a),(c),(e) The value of the parameter $\epsilon=0.01$. The PDAB solution is stable as observed in the intensity plot (a), and peak value of intensity (e). (b),(d),(f) The case when $\epsilon=0.05$. Intensity plot in (b) shows occurence of localizations at different locations with varying intensities.}
\label{figure7}
\end{framed}
\end{figure}

\begin{figure}
\centering
\begin{framed}
\begin{subfigure}{0.49\textwidth}
\centering
\includegraphics[width = 1 \textwidth]{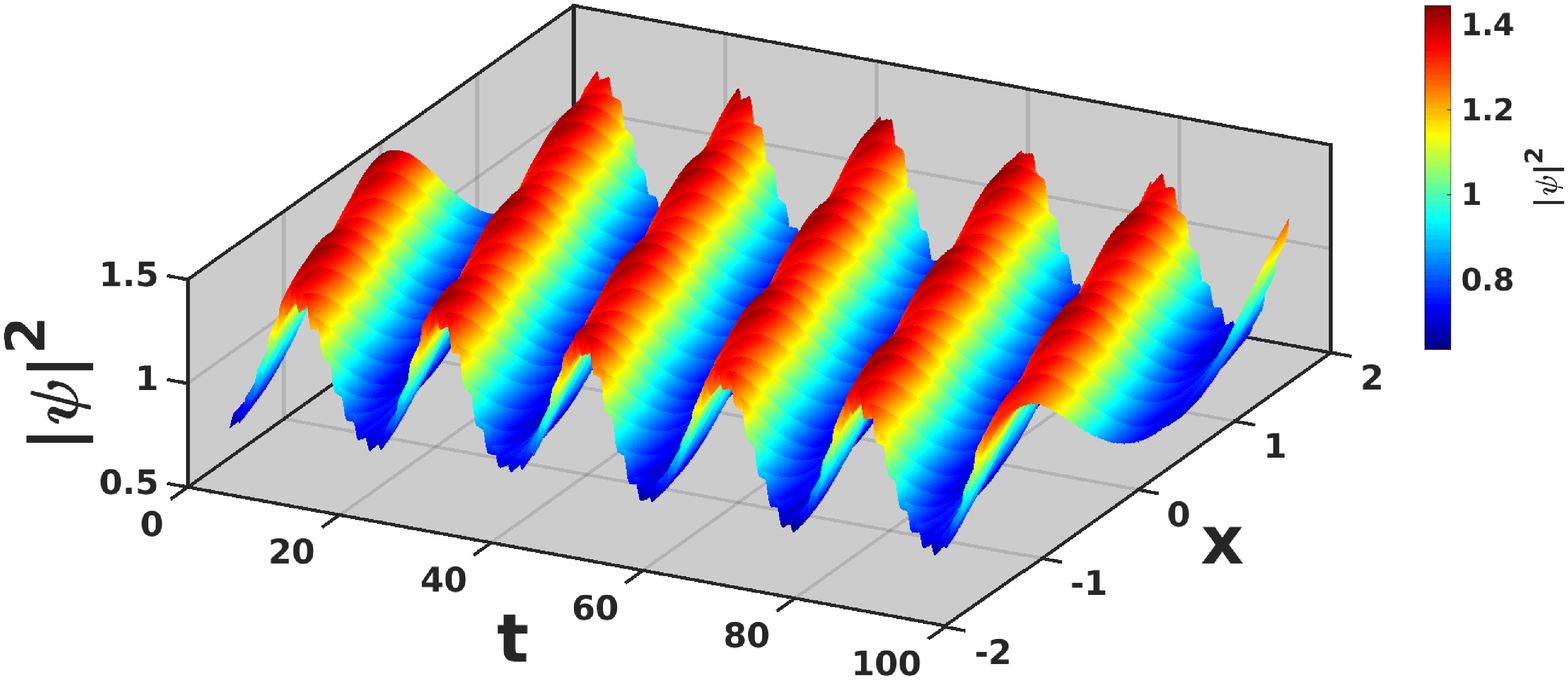}
\caption{}
\label{figure8a}
\end{subfigure}
\begin{subfigure}{0.49\textwidth}
\centering
\includegraphics[width = 1 \textwidth]{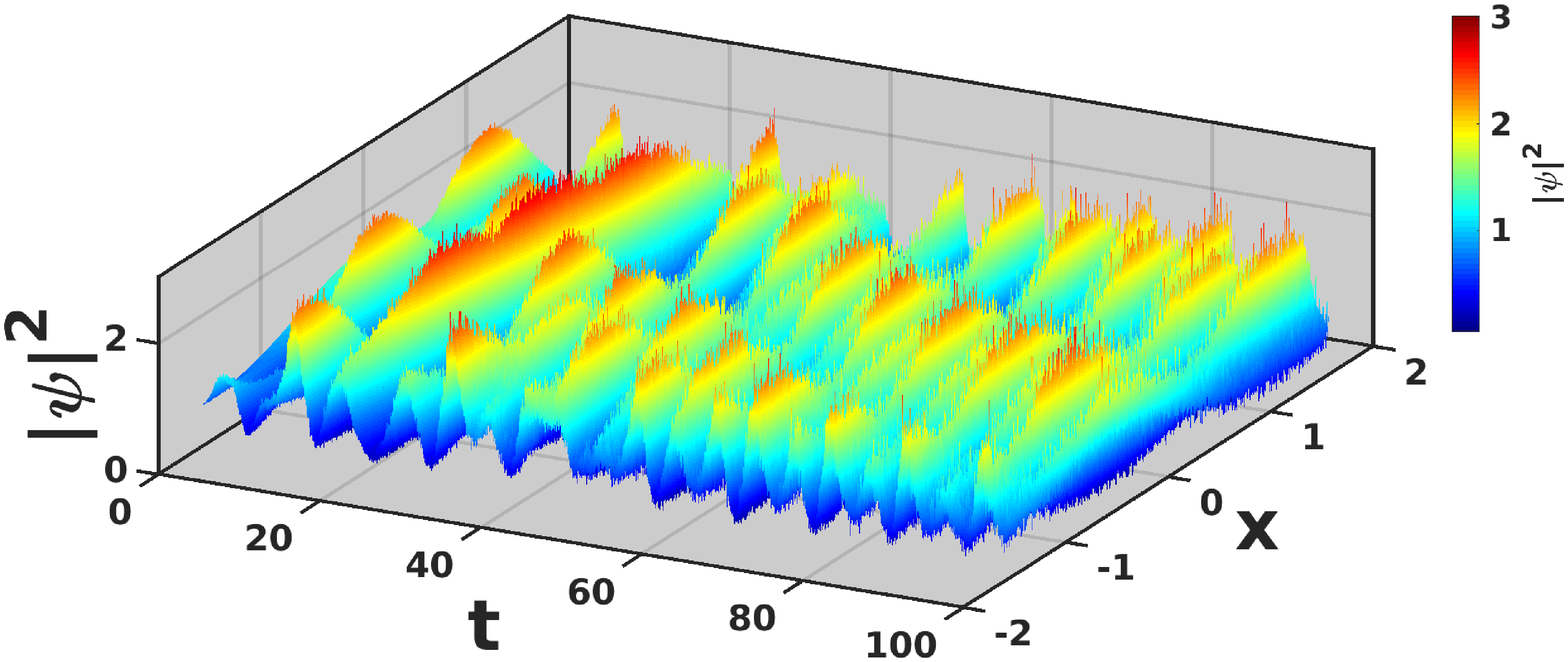}
\caption{}
\label{figure8b}
\end{subfigure}
\begin{subfigure}{0.49\textwidth}
\centering
\includegraphics[width = 1 \textwidth]{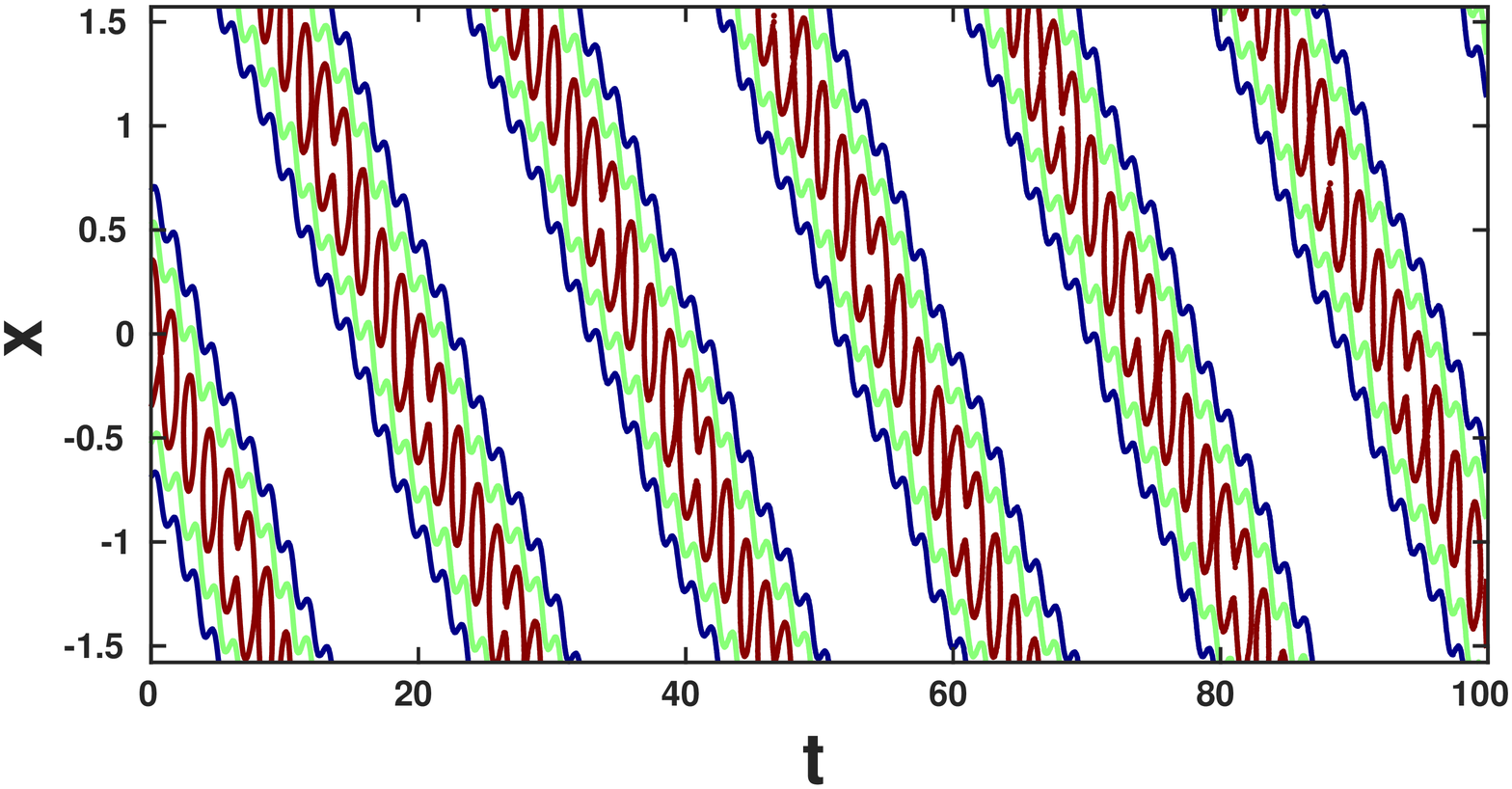}
\caption{}
\label{figure8c}
\end{subfigure}
\begin{subfigure}{0.49\textwidth}
\centering
\includegraphics[width = 1 \textwidth]{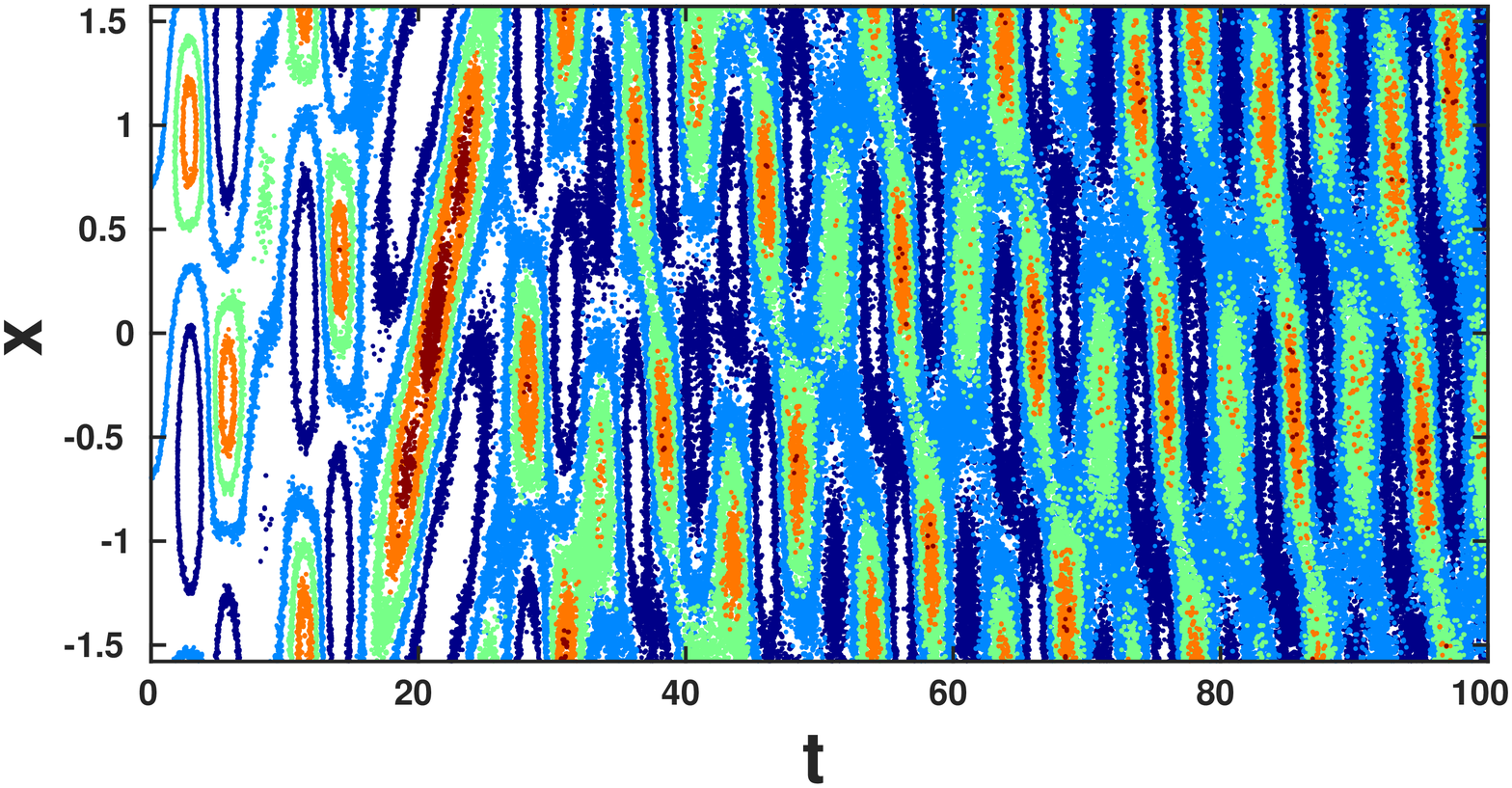}
\caption{}
\label{figure8d}
\end{subfigure}
\begin{subfigure}{0.49\textwidth}
\centering
\includegraphics[width =1\textwidth]{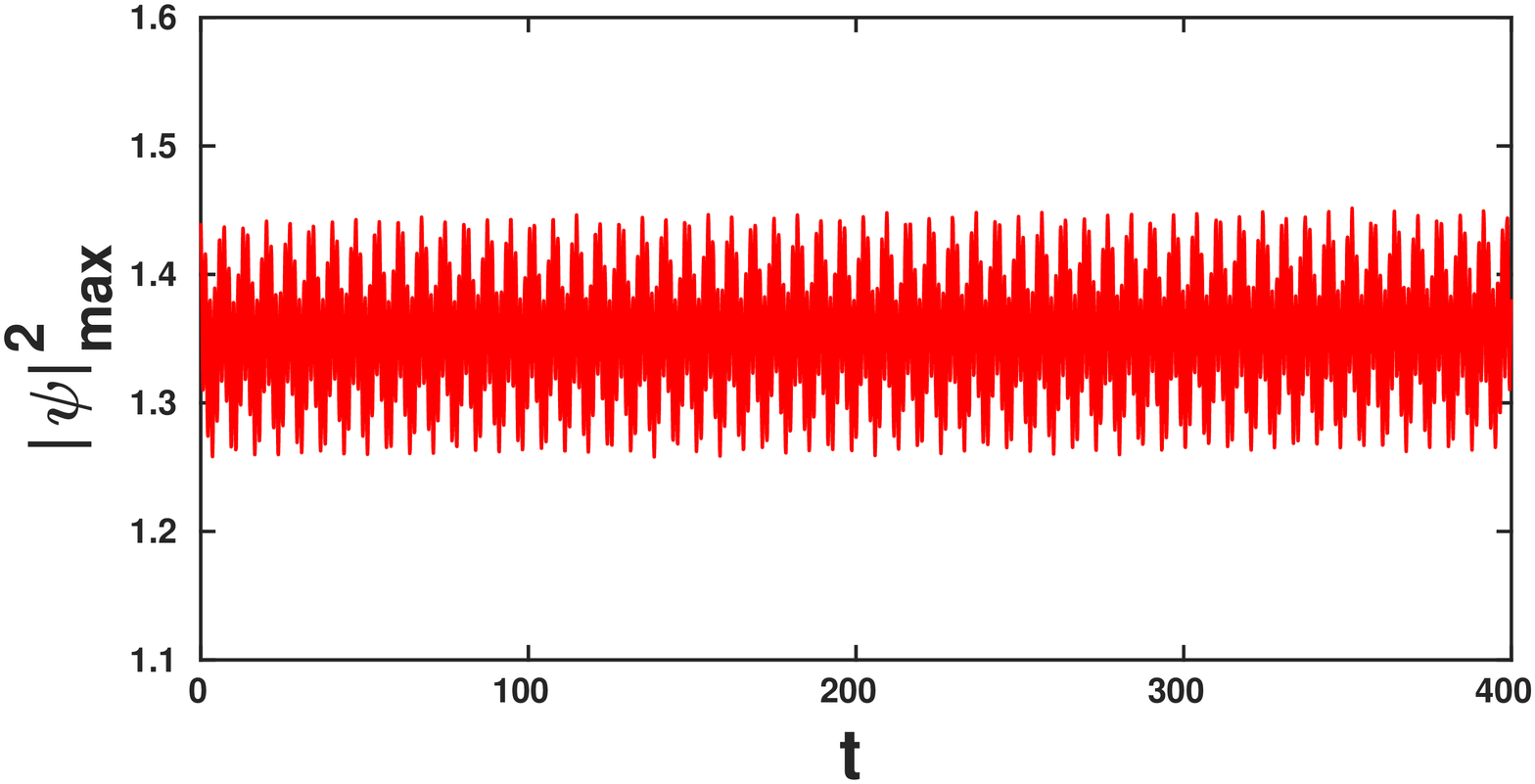}
\caption{}
\label{figure8e}
\end{subfigure}
\begin{subfigure}{0.49\textwidth}
\centering
\includegraphics[width =1\textwidth]{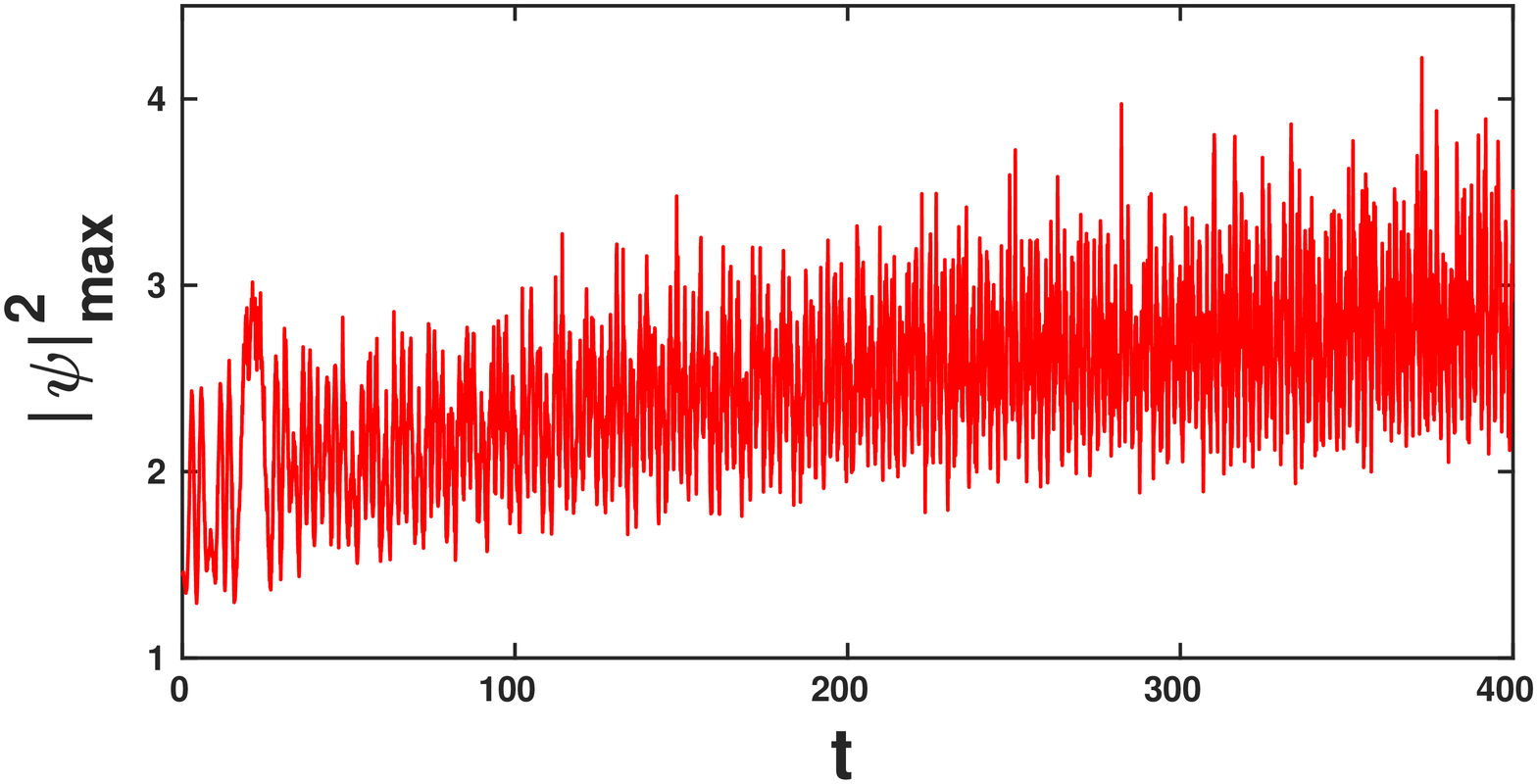}
\caption{}
\label{figure8f}
\end{subfigure}
\caption{Effect of multiplicative white noise on the PDAB solution. (a),(c),(e) The value of the parameter $\epsilon=0.05$. The PDAB solution is stable as observed in the intensity plot (a), and peak value of intensity (e). (b),(d),(f) The case when $\epsilon=5$. The solution is more stable in the presence of multiplicative noise.}
\label{figure8}
\end{framed}
\end{figure}

In reality, a dynamical system is always subjected to random perturbations. Such fluctuations could arise due to inhomogeneities in the system e.g. fluctuations in the refractive index of an optical medium~\cite{schwartz}. These perturbations, however small, can make a soliton unstable and therefore pose a challenge in using solitons for practical applications. Hence it is worthwhile to investigate the robustness of solitons against random perturbations. Accordingly, in this section, we study the stability of the PDAB solution in the presence of noise. In particular, we consider the evolution equation 
\begin{equation}
\label{eq10}
\centering
i\psi_t+\psi_{xx}+2|\psi|^2\psi=f_0  \e^{iKx}  \psi^*+\epsilon \ \mathcal{N}(\psi;x,t)
\end{equation}
where $\mathcal{N}(\psi;x,t)$ is a stochastic noise term and $\epsilon$ is a small parameter. We consider two types of noise as described in the following sections.
\subsection{Additive noise}
For the additive noise, the function
\begin{equation}
\label{eq11}
\centering
\mathcal{N}=\eta(x,t)
\end{equation}
such that 
\begin{equation}
\label{eq12}
\centering
\eqalign{\langle \eta(x,t) \rangle =0 \cr
\langle \eta(x,t)\eta(x',t') \rangle = \delta(t-t') \delta(x-x')} 
\end{equation}
where $\langle \cdots \rangle$ denotes ensemble average. The results obtained by using (\ref{eq11}) are shown in figure (\ref{figure7}). When $\epsilon = 0.01$, the PDAB is stable except for small local variations in the intensity (figure(\ref{figure7a})). As $\epsilon$ increases, fluctuations in the intensity also increases and when $\epsilon$ is comparable to the magnitude of the parametric forcing, the soliton structure is destroyed. Further increasing $\epsilon$ leads to isolated localizations in the intensity plot (figure(\ref{figure7b})). These intensity maxima are similar to breather excitaions, with a moderate increase in intensity (figure(\ref{figure7f})). Furthermore, we note that the recurrences do not exhibit any periodic behaviour.  

\subsection{Multiplicative noise}
Multiplicative noise, in the form
\begin{equation}
\label{eq13}
\mathcal{N}(\psi;x,t)=\eta(x,t) \psi(x,t)
\end{equation}
where $\eta(x,t)$ is as in (\ref{eq12}), is a type of noise that appears in physical situations such as in light propagation through an optical medium with random linear fluctuations in the refractive index, and  in the dynamics of Bose-Einstein condensate in disordered potentials~\cite{schwartz,schulte}. The effect of multiplicative noise on the PDAB is depicted in figure (\ref{figure8}). It is apparent from figure (\ref{figure8a}) that, for small values of noise, the intensity profile of the PDAB is almost indistinguishable from its noiseless dynamics. In fact, fluctuations in the maximum value of the intensity show similar pattern as in the zero noise case (figure(\ref{figure8e})). At higher values of noise, the soliton is replaced by breather-like excitations with aperiodic recurrence. Figure (\ref{figure8b}) shows the intensity plot when $\epsilon =5$. We note that, as in the additive case, the maximum value of the intensity increases with time (figure(\ref{figure8f})).

Although both additive and multiplicative noise eventually lead to instabilities in the soliton structure, the range of noise intensities for which the soliton remains stable, is different. Specifically, the PDAB is observed to be more stable against multiplicative random noise in comparison to the additive noise.

\section{Conclusion}
In conclusion, we have investigated the dynamics of Akhmediev breather under parametric driving. We have observed through numerical simulations that, for certain range of parameters of the system and initial conditions, the initial breather travels like a soliton whose amplitude and velocity are constants. The speed of the soliton is determined by the magnitude of forcing ($f_0$) and its direction is dictated by the sign of $f_0$. Although these new solutions are structurally similar to the conventional solitons, they are characteristically different from the ordinary solitons. Specifically, we notice that when dissipation is introduced, the amplitude of the soliton decreases while its velocity increases. We remark that, such an unconventional behaviour also opens up the possibility for new soliton solutions whose amplitude and velocity are inversely related. Furthermore, we have studied the stability of these solutions under random perturbations and found that the soliton is stable for sufficiently large values of perturbations.

\section*{References}
\bibliographystyle{iopart-num}
\bibliography{ref}
\end{document}